\documentclass[graybox]{svmult}

\usepackage{type1cm}        
\usepackage{makeidx}         
\usepackage{graphicx}        
\usepackage{multicol}        
\usepackage[bottom]{footmisc}

\usepackage{newtxtext}       %
\usepackage[varvw]{newtxmath}       

\makeindex             

\begin{document}

\title*{X-ray polarimetry-timing}
\author{Adam Ingram}
\institute{Adam Ingram \at School of Mathematics, Statistics and Physics, Newcastle University, Herschel Building, Newcastle upon Tyne, NE1 7RU, UK. \email{adam.ingram@newcastle.ac.uk}
}
%
%
\maketitle
\abstract{X-ray polarimetry-timing is the characterization of rapid variability in the X-ray polarization degree and angle. As for the case of spectral-timing, it provides causal information valuable for reconstructing indirect maps of the vicinity of compact objects. To call X-ray polarimetry-timing a young field is somewhat of an understatement, given that the first X-ray mission truly capable of enabling polarimetry-timing analyses has only just launched at the time of writing. Now is therefore an exciting time for the field, in which we have theoretical predictions and are eagerly awaiting data. This Chapter discusses the theoretical expectations and also describes the data analysis techniques that can be used.}

\section{Introduction}
\label{sec:intro}

We have already seen that rapid variability analysis of X-ray light curves provides an excellent means to study astrophysical compact objects -- black holes (BHs) and neutron stars (NSs). That these systems are compact means on the one hand that they cannot be spatially resolved, driving the \textit{need} for variability analysis for the purposes of indirect mapping, and on the other hand that they radiate a large X-ray luminosity that can be highly variable. The fact that they are compact therefore itself provides a solution to the problem that the compactness itself poses. Spectral-timing analysis -- i.e. characterising how the X-ray spectral shape varies on short timescales -- provides much more information than conducting separate X-ray timing and spectroscopy analyses, as it provides a probe of causality. For instance, two separate spectral components can be identified by X-ray spectroscopy, whereas a spectral-timing analysis can identify whether variations in one of those components lead or lag those in the other component. This causal information enables us to break degeneracies associated with spectral modelling and to make physical inferences. An extra dimension can be added to our arsenal if we are able to characterise how the X-ray \textit{polarization} varies on short timescales. In fact, such an \textit{X-ray polarimetry-timing} analysis can be seen as adding two extra dimensions, since we can measure both the polarization degree, $p$, and angle, $\psi$ (i.e. the extent to which the orientation of electric field oscillations all line up between different light rays, and the axis of that preferred orientation). Note that there is in principle a third extra dimension if circular polarization can be measured. However, there are no previous or planned X-ray polarimetry missions sensitive to circular polarization, and so for the remainder of this Chapter, `polarization' can be taken to exclusively mean `linear polarization'.

This Chapter first discusses scenarios in which we theoretically expect the polarization degree and angle to vary on short timescales. This covers essentially all varieties of compact objects: isolated NSs with periodic X-ray emission powered by rotational energy, accreting NSs and stellar-mass BHs in X-ray binary systems, and supermassive BHs in active galactic nuclei (AGNs). Attention is paid to how polarimetry-timing can help to break existing degeneracies and diagnose the physics associated with these objects. For instance, NSs are attractive sources because they inform on the equation of state of ultra-dense matter, and physics at the extremes of magnetic field strength. BHs enable us to probe the extremes of strong gravity, and the influence of AGNs on galaxy formation. Also of great interest are the formation and evolution channels of the objects themselves, which can be constrained from the observed distribution of BH mass and spin.

Polarimetry-timing is a very young field, since at the time of this Chapter being written a period of $\sim 40$ years without a space-based polarimeter being in operation is finally drawing to a close. A number of X-ray polarimeters were launched in the 1970s, the last of which being OSO-8. These early polarimeters worked via Bragg reflection, such that polarization revealed itself as a modulation of the X-ray count rate at half the $10$ s rotation period of the polarimeter assembly around the reflector. The effective area was small, and so sensitivity was low. OSO-8 measured the X-ray polarization of the Crab nebula with high statistical significance \cite{Weisskopf1978}, but only achieved upper limits for other targets such as the NS X-ray binary Scorpius X-1 \cite{Weisskopf1978a} and the BH X-ray binary Cygnus X-1 \cite{Long1980}. Notably, the first polarimetry-timing experiment was performed with OSO-8: Silver et al \cite{Silver1978} attempted to isolate the X-ray polarization of the Crab pulsar from the surrounding nebula with pulse-phase resolved polarimetry. They created two light curves, one corresponding to pulse-phases when the pulsed emission was at a minimum and the other corresponding to the remaining phases. The former light curve was taken to be dominated by the nebula. They then separately folded these two light curves on the instrument rotation period to measure the polarization, and finally subtracted one from the other to determine if the polarization of the pulsar itself could be constrained. Only upper limits were obtained, but this analysis was the first example in the literature of any sort of X-ray polarimetry-timing analysis. After OSO-8, more than $40$ years passed without the launch of another space-based observatory sensitive to X-ray polarisation. Very recently, balloon borne experiments have recorded interesting upper limits \cite{Chauvin2018} and, most notably, the cubesat PolarLight detected polarisation in the 4-8 keV band from Scorpius X-1 \cite{Long2022}, confirming the earlier marginal OSO-8 detection. The true breakthrough, however, came with the launch of the Imaging X-ray Polarimetry Explorer (IXPE) on $9^{\rm th}$ December 2021. As perhaps expected from the 40 year time interval between the missions, IXPE is far more sensitive than OSO-8. The step up in sensitivity is a factor $\sim 100$, mainly driven by the much larger effective area (this is discussed in much more detail elsewhere in this book). Not only does this increase in sensitivity promise to enable the first statistically significant measurements of polarization for a number of source classes such as BH X-ray binaries and AGN, it also enables polarimetry-timing analyses beyond the solitary $>40$ year old example of the Crab pulsar investigation. Planned future missions, such as the enhanced X-ray Timing and Polarimetry mission (eXTP), have the potential to further push this new field forward.

There is, however, a technical challenge. It takes many photons to measure X-ray polarization with statistical significance, and therefore a long exposure. If we wish to detect variability in polarization properties by directly measuring $p$ and $\psi$ for a series of contiguous time bins, we would typically need the duration of these time bins to be longer than the variability timescales of interest. More sophisticated techniques than naively extracting a light curve of $p$ and $\psi$ are therefore required. For periodic signals such as pulsations, the solution is clear. One can simply phase-resolve on the known pulse period as Silver et al \cite{Silver1978} did. However, for stochastic signals, phase-resolving is not possible because the phase does not increase predictably with time. This Chapter discusses the optimal techniques to enable polarimetry-timing analyses of such stochastic signals.

\section{Theoretical expectations}
\label{sec:theory}

This Section discusses different kinds of variability, from a variety of astrophysical objects, that we may theoretically expect in the X-ray polarization signal.

\subsection{Pulsations}
\label{sec:pulsars}

NSs exhibit X-ray pulsations in a number of scenarios, typically due to regions of enhanced emission (`hotspots') at the magnetic poles rotating with the star due to a misalignment between the magnetic and rotation axes \cite{Pechenick1983,Poutanen2003}. For NSs in binary systems, the hotspots are either accretion powered \cite{Wijnands1998} or form in the aftermath of a thermonuclear burst (burst oscillations \cite{Strohmayer2006,Watts2012}). For isolated systems, the hotspots are caused by magnetospheric currents depositing energy in the surface layers of the star \cite{Ruderman1975}. The energy-dependent pulse profile depends on the geometry of the system: the size of the hotspot, the misalignment angle between the magnetic pole and NS rotation axis $\theta$, and the observer's inclination angle $i$ (the angle between the NS rotation axis and the observer's line of sight). Crucially, the pulse profile is distorted by relativistic effects, enabling inference of the NS mass $M$ and radius $R$, which in turn constrain the equation of state of ultra-dense matter \cite{Watts2019}. Light-bending, aberration, lensing and gravitational redshift depend only on the compactness $R/M$, whereas Doppler boosting due to NS rotation depends on the rotation velocity at the NS surface. This velocity depends on both $R$ and the rotation frequency of the NS, but the later is very simply measured from the pulse period. The mass and radius can therefore both be constrained, but a reasonably high spin frequency ($\nu \gtrsim 100$ Hz) is required for rotational Doppler boosting to have a strong enough effect on the pulse profile.

\begin{figure}[b]
\includegraphics[scale=.6]{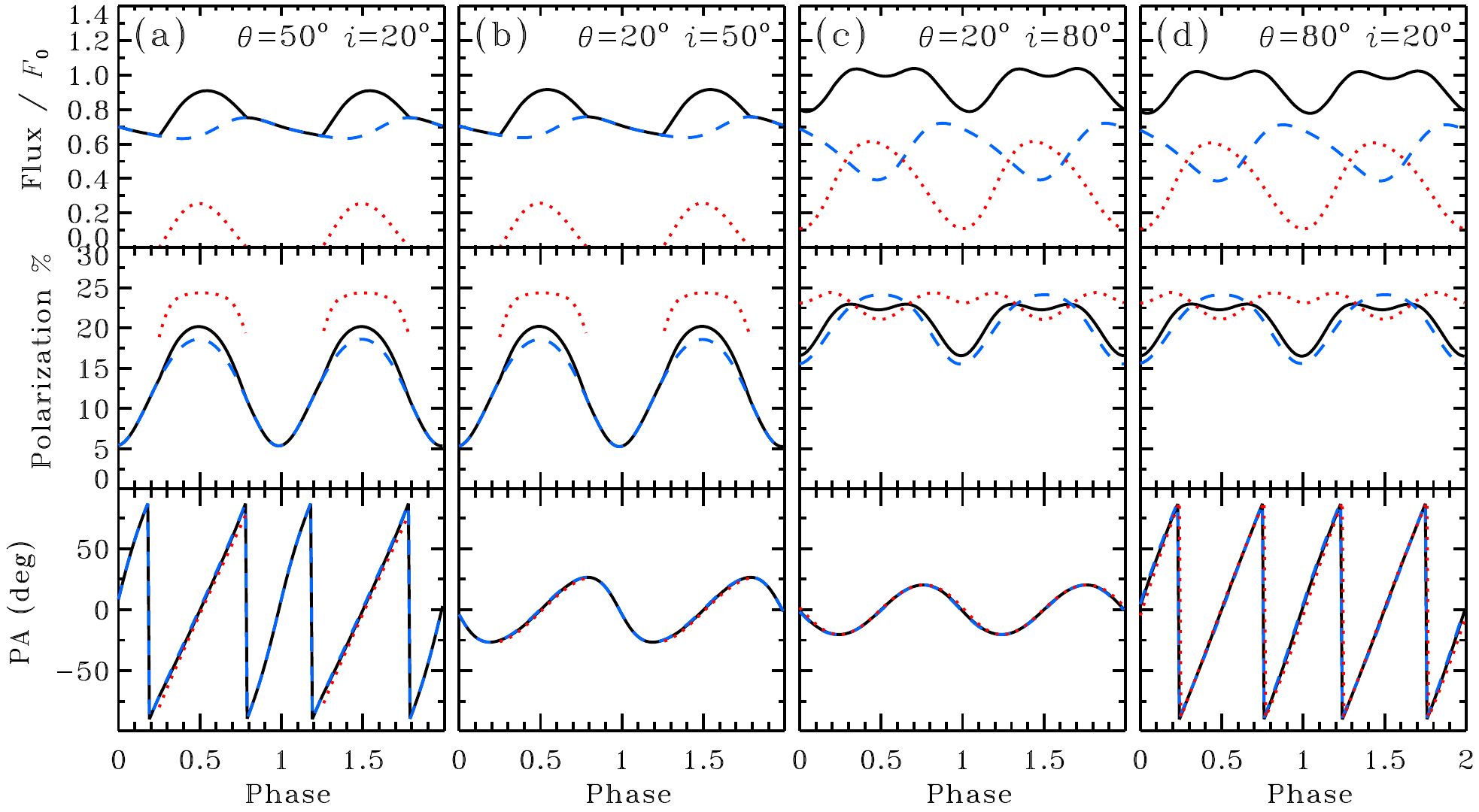}
\caption{Pulse profile models from \cite{Viironen2004}, figure reproduced from \cite{Watts2019}. From top to bottom, the X-ray flux, polarization degree and polarization angle are plotted against pulse phase for different combinations of the geometric parameters $\theta$ and $i$. Other parameters are: NS spin $\nu=300$ Hz, $M = 1.4~M_\odot$ and $R=5~GM/c^2\approx 10.3$ km. The contribution from the primary (blue dashed) and secondary hotspots (red dotted) are plotted separately, and the solid black line is the sum of the two. The flux and polarization degree are almost completely degenerate to exchanging $\theta$ and $i$, whereas the two alternative geometries are easily differentiated by the polarization angle.}
\label{fig:pulse}       
\end{figure}

Recently, $\sim 1.9$ Ms of exposure with NICER was used to infer $M=1.34^{+0.15}_{-0.16}~M_\odot$ and $R=12.71^{+1.14}_{-1.19}$ km for the $\nu=205$ Hz isolated pulsar PSR J0030+0451 \cite{Riley2019}. Such a large exposure is required to overcome parameter degeneracies in the modelling procedure, which prevented strong constraints from being achieved with RXTE data \cite{Poutanen2003}. Fig. \ref{fig:pulse} demonstrates one such parameter degeneracy, using models from \cite{Viironen2004}. The black solid lines represent the flux summed over both hotspots and the blue dashed and red dotted lines represent flux from the primary and secondary hotspot respectively. The hotspots are assumed to be antipodal, but note that there is strong evidence that this is not the case for PSR J0030+0451 \cite{Riley2019}. The top row shows X-ray flux. We see that this is almost completely degenerate to a switch between the angles $\theta$ and $i$. The second and third rows show polarization degree and angle respectively. We see that polarization degree is also degenerate to switching the angles $\theta$ and $i$, but the polarisation angle is very much not. Therefore measuring the pulse phase dependent polarisation angle will easily break a key degeneracy of the pulse profile modelling technique.

In the simplest model for the polarisation angle swings (the rotating vector model \cite{Radhakrishnan1969}), the polarisation vector is simply the projection of the magnetic polar axis on the image plane such that the polarisation angle, $\psi$, is given by
\begin{equation}
    \tan\psi = - \frac{\sin\theta\sin\phi}{\sin i \cos\theta - \cos i \sin\theta \cos\phi},
\end{equation}
where $\phi$ is pulse phase. Relativistic effects such as light bending and parallel transport in a warped spacetime adjust the polarisation angle from this simple expression, but the above gives an intuitive approximation.

\begin{figure}[b]
\includegraphics[width=\textwidth,trim=0.0cm 20.5cm 0.0cm 0.0cm,clip=true]{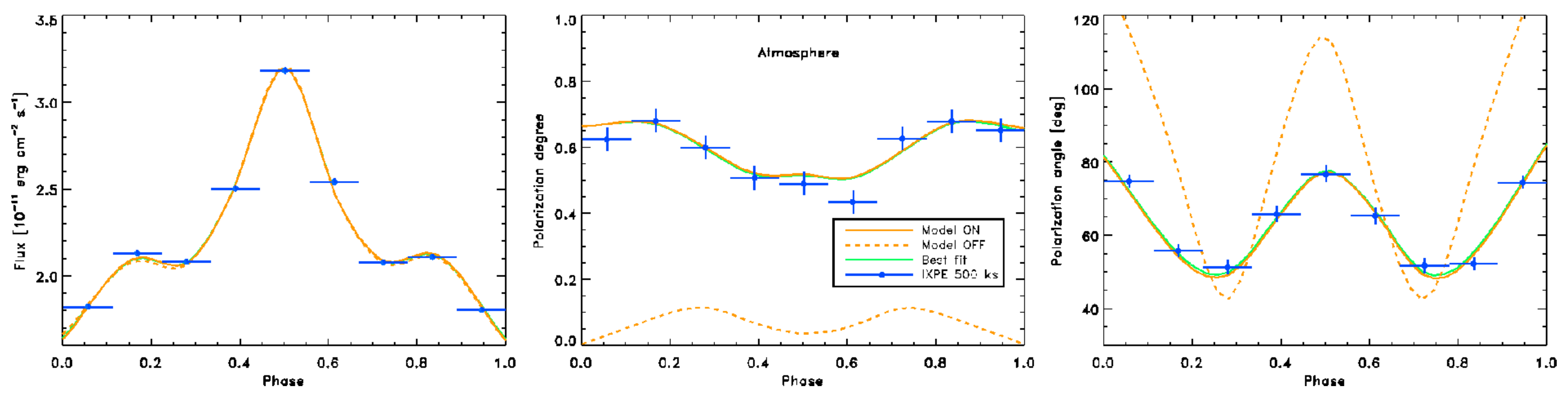}
\caption{Pulse phase dependence of X-ray flux (left), polarization degree (middle) and polarization angle (right) for persistent (low luminosity) magnetar emission predicted by the model of \cite{Taverna2020}. Here, $i=90^\circ$ and $\theta=60^\circ$, the magnetic field strength at the poles is $B=5\times 10^{14}$ G, and other parameters are appropriate for the magnetar 1RXS J170849.0-400910. The solid orange lines represent the full model and the dashed orange lines represent the same model with the QED vacuum birefringence effect artificially switched off. This makes essentially no difference to the flux but dramatically changes the polarization properties. The blue points represent a simulated 500 ks IXPE observation, and the solid green line represents the best fit model to those synthetic data.}
\label{fig:qed}       
\end{figure}

The NSs with the strongest magnetic fields provide the opportunity to test other aspects of fundamental physics. Magnetars have an inferred dipole magnetic field strength of $B \gtrsim 10^{14}$ G. At such high field strengths, the theory of quantum electrodynamics (QED) predicts that vacuum birefringence will have a significant effect on the observed X-ray polarisation \cite{Heyl2000,Heyl2002,Taverna2014}. This effect of Faraday rotation during propagation of photons through a vacuum is a fundamental prediction of QED that has never been verified, since it is not possible to produce a strong enough magnetic field in a terrestrial laboratory. Another predicted effect is magnetic condensation of the NS atmosphere, which occurs at some critical field strength that increases with temperature \cite{Lai2001}. Observational estimates indicate that the known population of magnetars includes objects with field strength both above and below this critical value \cite{Taverna2020}.

Fig. \ref{fig:qed} shows predicted pulse profiles for a magnetar with polar magnetic field strength $B=5\times 10^{14}$ G, adapted from \cite{Taverna2020} (other parameters in the caption). The solid orange lines represent the full model that includes QED effects, whereas for the dotted orange lines QED has been artificially switched off. We see that the flux pulse profile (left) is completely unaffected by vacuum birefringence, whereas QED has a dramatic effect on the polarization degree (centre) and angle (right). Measuring the pulse phase-resolved X-ray polarization degree and angle will therefore provide a new means to verify the predictions of QED. The blue points represent a simulated $500$ ks IXPE exposure, demonstrating that the required accuracy for QED tests is achievable in the near future. The phase-resolved polarization degree and angle also depend on whether the B-field strength is above or below the critical value for magnetic condensation \cite{Taverna2020}, and so this can in principle be determined with future X-ray polarimetric observations.

\subsection{Propagating accretion rate fluctuations}
\label{sec:propfluc}

All accreting BHs and NSs exhibit aperiodic variability in their X-ray flux, with properties that evolve as the X-ray spectrum evolves from the power-law dominated hard state to the thermal dominated soft state via the intermediate states \cite{VDK2006,Belloni2010}. For BH X-ray binaries, the power spectrum in the hard state is roughly white ($P\propto \nu^0$) below a low frequency break $\nu_{\rm lo}$, steeper above this break $P\propto \nu^{-1}$, and steeper still ($P\propto \nu^{-2}$) above a high frequency break $\nu_{\rm hi}$. This means that there is strong variability on a range of timescales between $1/\nu_{\rm hi}$ and $1/\nu_{\rm lo}$, with little variability on timescales shorter than $1/\nu_{\rm hi}$ or longer than $1/\nu_{\rm lo}$. As the source evolves through the hard state and towards the soft state, all characteristic frequencies in the power spectrum increase \cite{Wijnands1999,Psaltis1999}. AGN power spectra are consistent with being mass and accretion rate scaled versions of BH X-ray binary power spectra \cite{McHardy2006}. NS X-ray binary power spectra display a similar (mass-scaled) phenomenology but are typically more complex, perhaps due to processes associated with the solid stellar surface \cite{vanstraaten2002}. The variability is characterised by the linear rms-flux relation \cite{Uttley2001}: epochs of high X-ray flux have a systematically larger variability amplitude such that splitting up a long light curve into many segments and calculating the mean flux and standard deviation for each segment will reveal that the two are linearly related (after binning in flux). The relation holds for any set of timescales (i.e. light curves with any time bin and segment length). This rules out simple shot noise models whereby the light curve is reproduced by a superposition of different, unrelated flares each with duration randomly drawn from a distribution, instead indicating a causal connection between variability of different characteristic time scales \cite{Uttley2005}. The relation also holds for all accreting sources, including BHs, NSs, white dwarfs and even young stellar objects \cite{Scaringi2015}. Another key property of the aperiodic variability is the Fourier frequency dependent time lags. Hard photons are observed to lag soft photons for low Fourier frequencies ($\nu \lesssim 100~M_{\odot}/M_{\rm bh}$ Hz). The time lag reduces with increasing frequency, (roughly $t_{\rm lag} \propto \nu^{-0.7}$ \cite{Nowak1999}), and increases with the energy difference between the two considered light curves (i.e. $t_{\rm lag} \propto \log(E_2/E_1)$, where the lag is measured between energy bands centered at $E_1$ and $E_2$ \cite{Kotov2001}).

The above phenomenology can be explained by the propagating accretion rate fluctuations model \cite{Lyubarskii1997,Kotov2001,Arevalo2006,Ingram2016c}. Variability in the accretion rate is assumed to be stirred up throughout the entire accretion flow by magnetic turbulence (the magneto-rotational instability \cite{Balbus1991}), with the characteristic variability timescale related to the local viscous timescale. The longer timescale variability is therefore generated further from the BH, and the faster variability is generated in the inner regions. As material drifts towards the BH, accretion rate fluctuations propagate inwards, modulating the faster variability produced at smaller radii. This means that fluctuations with different characteristic timescales combine \textit{multiplicatively}, naturally reproducing the linear rms-flux relation. The time lags can also be in principle explained if the accretion flow emits a progressively harder spectrum from smaller radii. Thus an accretion rate fluctuation far from the BH causes first a flare in the soft X-rays, and then after a propagation time another flare in the hard X-rays. The frequency and energy dependence of the time lags are also naturally understood in this picture, since slow fluctuations can originate anywhere in the accretion flow, whereas the fastest fluctuations can only originate close to the BH. Therefore, on average, the time lags at high frequency are shorter than those at low frequency, because the propagating fluctuations at high frequencies were all generated fairly close to the BH and therefore did not have far to travel before reaching the hard X-ray emitting region.

\begin{figure}[b]
\includegraphics[width=0.5\textwidth,trim=1.0cm 1.0cm 7.0cm 4.0cm,clip=true]{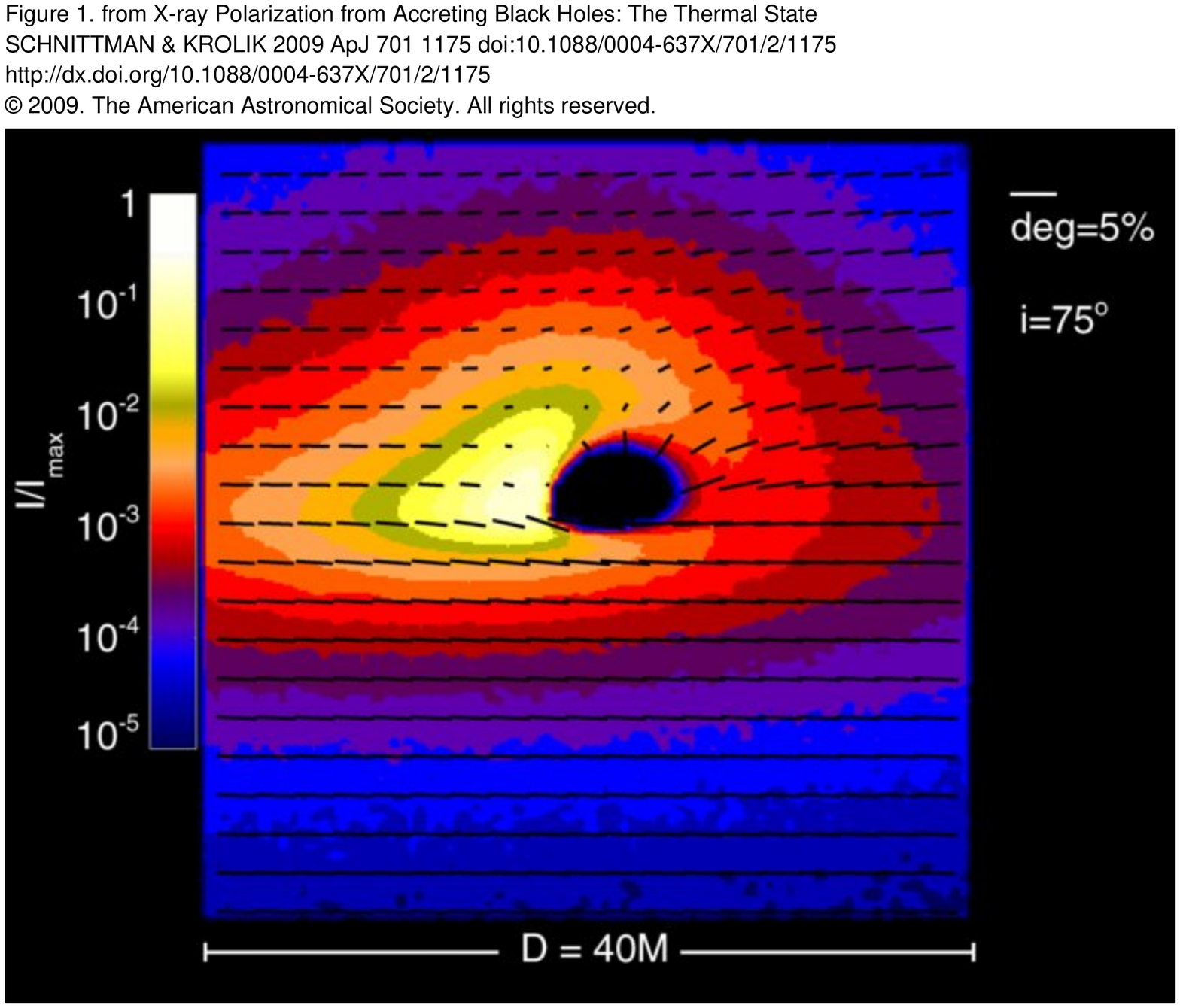}
\includegraphics[width=0.5\textwidth,trim=1.0cm 1.0cm 7.0cm 4.0cm,clip=true]{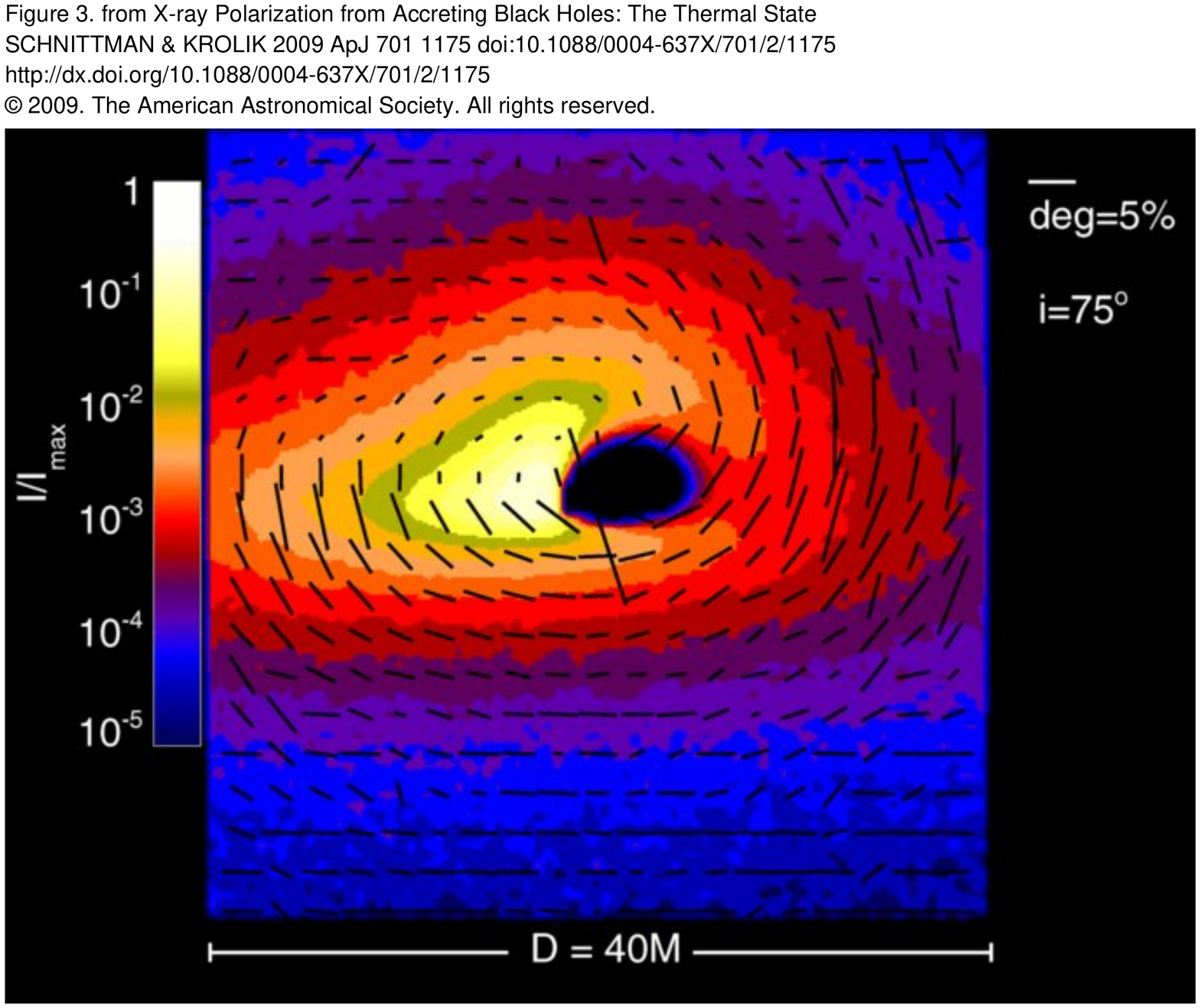}
\caption{Maps of flux (colour scale), polarization degree (represented by the length of the black holes) and polarization angle (represented by the orientation of the black lines) from a standard Novikov \& Thorne \cite{Novikov1973} accretion disc around a BH with dimensionless spin parameter $a=0.99$. Reproduced from \cite{Schnittman2009}. The left hand plot only includes direct radiation, whereas the right hand plot also includes returning radiation. The returning photons have reflected from the disc at least once after their initial emission and are therefore highly polarized.}
\label{fig:discpol}       
\end{figure}

\begin{figure}[b]
\sidecaption
\includegraphics[width=\textwidth,trim=0.0cm 6.0cm 0.0cm 0.0cm,clip=true]{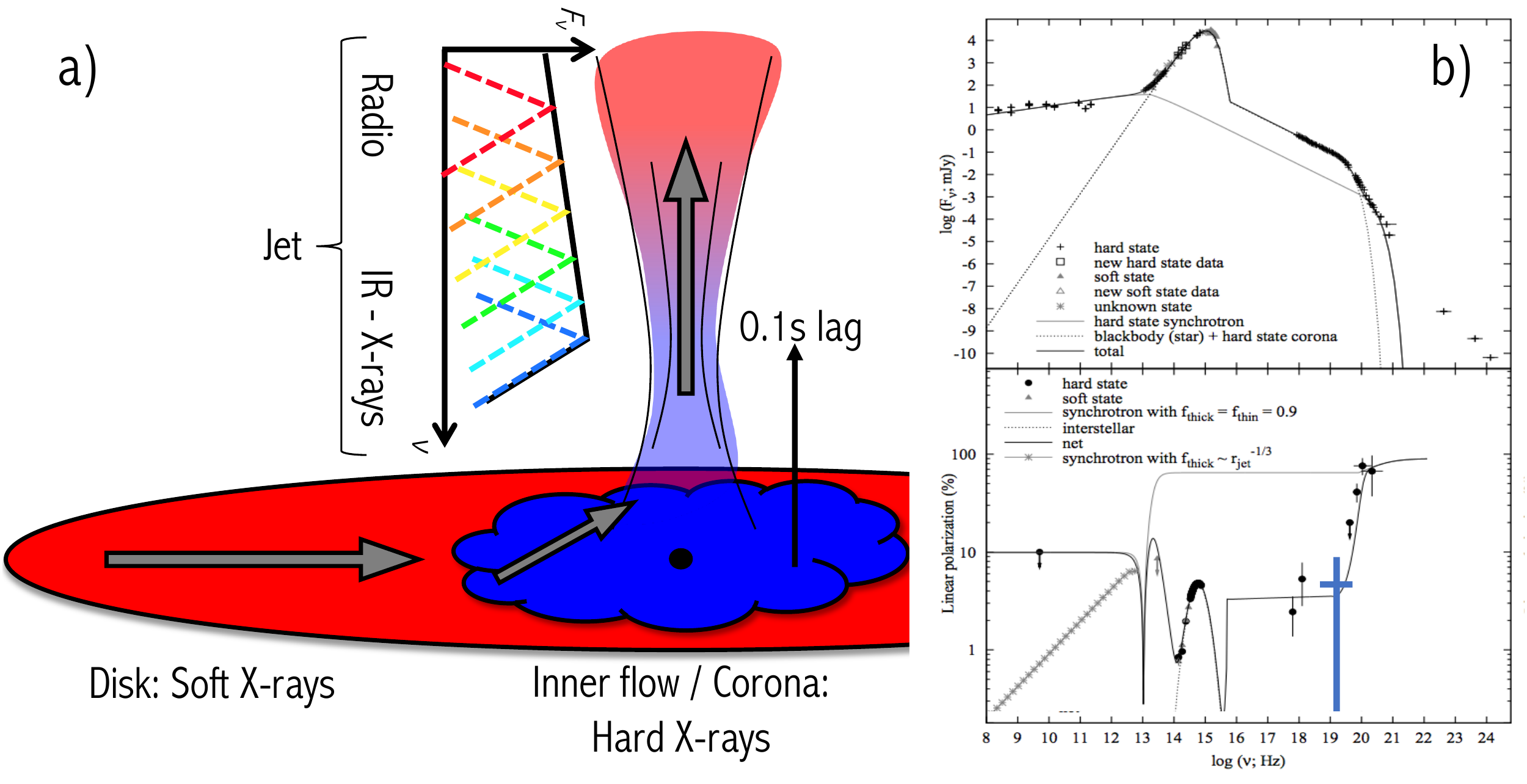}
\caption{\textit{a):} Schematic of accretion rate fluctuations propagating through the disc, then corona and up the jet. The jet spectrum is pictured as a superposition of self-absorbed synchrotron components, and so is optically thin above a high frequency break. \textit{b):} The broad band spectrum of Cygnus X-1 fit with a simple toy model (adapted from \cite{Russell2014}). The companion star and corona are assumed to be unpolarized, whereas the jet has moderate polarization below the break and high polarization above the break. The toy model reproduces the observed increase in polarization degree from soft X-rays to soft $\gamma-$rays, implying that the jet contributes $\sim 5\%$ of the soft X-ray flux. The blue hard X-ray point has been added for this figure, and is also consistent with the toy model.}
\label{fig:jet}       
\end{figure}

If the polarization degree and angle also depend on radius in the accretion flow, there should also be a signature of propagating accretion rate fluctuations in the time-dependent polarization properties. At the most basic level, a fluctuation propagating from a weakly polarized region to a highly polarized region will result in a highly polarized population of photons lagging behind a weakly polarized population. Or vice versa if propagation is from a highly polarized to a weakly polarized region. So do we expect the polarization degree and angle to depend on radius in an accretion disc? Parallel transport of vectors in a warped spacetime has the effect of rotating the polarization angle of a photon that travelled from close to the BH to an observer at infinity. We can think of this as `gravitational Faraday rotation'. This effect is far stronger for photons that pass closer to the BH and so there is a strong radial dependence, which is demonstrated in Fig. \ref{fig:discpol} (left) in which the entire disc is assumed to have the same polarization degree and angle locally. The result is that radiation from close to the BH has a lower overall polarization, because the radiation from different azimuths has a range of polarization angles leading to a cancelling of polarization in the summed emission. Since for a standard thin accretion disc, a harder spectrum is emitted from progressively closer to the BH, this leads to a swing in polarization angle and a drop in polarization degree above a characteristic energy \cite{Stark1977}, which can in principle be used to measure the BH spin from soft state observations \cite{Dovciak2008}. An extra effect that turns out to be very important is returning radiation: rays that, after being emitted from the accretion disc, were bent so dramatically by gravity that they were again incident on the disc before being reflected and travelling to the observer. Such reflected rays are highly polarized. Fig. \ref{fig:discpol} (right) includes this effect of returning radiation.

The above discussion is relevant to a thin, thermalised accretion disc, which is thought to be present in the soft state. However, the variability amplitude is very low during the soft state. If we wish to conduct polarimetry-timing investigations, they must be of the hard and intermediate states, which exhibit high levels of variability. The significant complication in these states is the presence of the X-ray corona. Compton scattering of photons in the corona will imprint a polarization signature on the X-ray radiation. Put simply, the more asymmetric the corona is, the more highly polarized its radiation field is. For example, a perfectly spherical corona with seed photons all originating from its centre will produce completely un-polarized radiation. If the seed photons instead come from the equatorial plane (i.e. the accretion disc), there will be a small but non-zero polarization, and there will be a higher polarization still if the corona is highly radially or vertically extended \cite{Sunyaev1985,Schnittman2010,Tamborra2018,Zhang2019}. Therefore, whether we expect high polarization photons to lag or lead low polarization photons is highly dependent on the model adopted for the corona. Polarimetry-timing therefore in principle presents an important extra diagnostic to constrain the coronal geometry. Currently, we only have spectral and timing properties, whereas soon we will also have energy and time dependent polarization properties to additionally test models against. However, the polarization of both the disc \cite{Chandrasekhar1960} and corona \cite{Sunyaev1985} is expected to be reasonably low ($\lesssim 10\%$), an expectation consistent with the small amount of observational data currently available to us \cite{Long1980,Chauvin2018}. The propagation lag between disc and corona is also expected to be small. Therefore the observational signal of rapid propagation from one weakly polarized region to another even more weakly polarized region will be very small indeed and practically not possible to detect with at least the next few generations of X-ray polarimeters.
 
It may, however, be possible to make a detection of propagation across a fairly large distance between two regions with very different polarization degrees. This could indeed be the case if fluctuations propagate through the disc, then the corona, and finally up the jet \cite{DeRosa2019}. The jet spectrum is thought to be a superposition of self-absorbed synchrotron components \cite{Blandford1979}. As is illustrated in Fig. \ref{fig:jet}a, the self-absorption frequency, at which each component peaks, reduces with increasing distance from the BH. The highest frequency component originates from the region where electrons are first accelerated to relativistic velocity (radius $r_{\rm acc}$). The total spectrum is therefore approximately flat below the highest self-absorption frequency, with a break to optically thin synchrotron emission at a frequency typically measured to be in the IR band \cite{Gandhi2011,Russell2014a}. Fluctuations in the optical and IR bands have been measured to lag behind those in the X-ray band by $\sim 0.1$ s for a few X-ray binaries \cite{Casella2010,Gandhi2017}. A picture whereby accretion rate fluctuations propagate through the corona and up the jet, taking $\sim 0.1$ s to reach $r_{\rm acc}$, is therefore consistent with the observations (Fig. \ref{fig:jet}a). Since optically thin synchrotron radiation is characterised by a high polarization degree (up to $\sim 70\%$ \cite{Rybicki1979}) if the magnetic fields are sufficiently ordered, the contribution to the spectrum from the jet could be highly polarized from the IR all the way up to soft $\gamma-$rays. If this optically thin component contributes $\sim 5-10 \%$ of the flux in the soft X-ray band and is $\sim 70\%$ polarized, it could therefore dominate the polarization signal over the weakly polarized corona that dominates the flux. Since the polarization difference and the time lag between the two components are both reasonably large, it may be possible to make a detection of the resulting time lags with IXPE. Just how to technically make such a measurement is discussed in the following Section.

Such a configuration with the soft X-ray polarization being dominated by a low flux but highly polarized optically thin synchrotron component has been suggested for Cygnus X-1. This is motivated by the high polarization degrees measured by the SPI and IBIS instruments on-board INTEGRAL of $76 \pm 15\%$ \cite{Jourdain2012} and $67\pm 30\%$ \cite{Laurent2011} respectively in the $>230$ keV and $>400$ keV energy bands (the count rate drops to approximately zero above $\sim 850$ keV). Fig. \ref{fig:jet}b shows the broadband spectrum of Cygnus X-1 with a very simple toy model consisting of an unpolarized companion star, an unpolarized corona and a synchrotron jet component that switches from a moderate polarization in the optically thick regime, to high polarization in the optically thin regime (photon frequency above $\sim 10^{13}$ Hz) \cite{Russell2014}. This toy model reproduces the broadband polarization degree remarkably well. The jet component dominates the soft $\gamma-$rays, leading to a high polarization, whereas the overall polarization is highly diluted by the weakly polarized corona, which contributes $\sim 95\%$ of the flux, in the soft band. The soft X-ray ($\sim 10^{18}$ Hz) data points here are from OSO-8 \cite{Long1980}, and the blue hard X-ray point from POGO+ \cite{Chauvin2018} was recorded after the original version of this plot was published. All three of these X-ray points are consistent with zero polarization within the $3\sigma$ confidence interval. On the face of it, this model is very appealing for Cygnus X-1. However, the X-ray polarization in the model aligns with the imaged direction of the radio jet
\cite{Miller-Jones2021}, requiring the magnetic field to be highly ordered perpendicular to the jet direction.
This requires a toroidal magnetic field configuration, which can only be strongly ordered when viewed edge-on (imagine viewing a tightly coiled spring edge-on -- you see the spring as a series of parallel lines). Cygnus X-1, in contrast, is viewed from a reasonably low inclination ($\sim 30^\circ$; \cite{Miller-Jones2021}), in which case we view the toroidal field lines as ovals. The polarization is still perpendicular to the jet direction, but the degree is far below $\sim 70\%$. It is therefore highly unlikely that the soft X-ray polarization is dominated by the jet base for the case of Cygnus X-1, but it is still possible for highly inclined objects, or objects whose jets have a poloidal magnetic field configuration.

\subsection{X-ray reverberation mapping}
\label{sec:reverb}

\begin{figure}[b]
\sidecaption
\includegraphics[width=\textwidth,trim=0.0cm 14.0cm 1.0cm 0.5cm,clip=true]{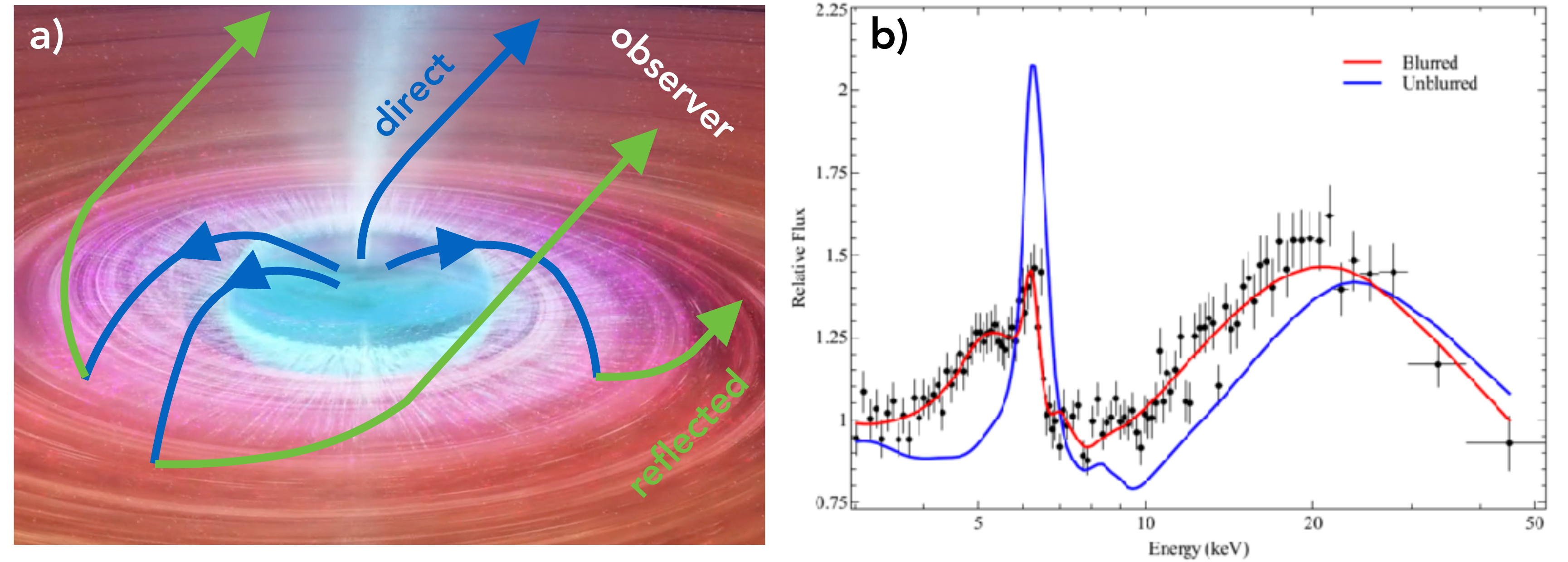}
\caption{\textit{a):} Schematic of reflection. Some X-rays radiated by the corona (blue) are directly observed and some irradiate the disc to be re-processed and re-emitted (green). \textit{b):} X-ray spectrum observed from the AGN Mrk 335 \cite{Parker2014}. Black points are NuSTAR data divided through by the best fitting direct continuum model and the red line shows the best fitting relativistically smeared reflection model. The blue line is the same reflection model in the disc restframe, plotted to illustrate the relativistic effects that are imprinted onto the spectrum. Adapted from \cite{Fabian2016}.}
\label{fig:reflection}       
\end{figure}

Whenever an X-ray corona is present, it will irradiate the disc (Fig \ref{fig:reflection}a). This radiation is re-processed in the disc atmosphere and re-emitted with a characteristic \textit{reflection} spectrum that includes features such as a complex of iron K emission lines at $\sim 6.4-6.9$ keV, and iron K edge at $\sim 7$ keV and a broad Compton scattering feature peaking at $\sim 20-30$ keV known as the Compton hump (Fig \ref{fig:reflection}b). The shape of this reflection spectrum is distorted according to the observer by energy shifts due to the rapid orbital motion of the disc material and gravitational redshift. In particular, the iron line is narrow in the disc restframe, but asymmetrically broadened in the observer frame such that the observed iron line profile can be used to infer, amongst other things, the disc inclination angle and inner radius in units of $r_g$ \cite{Fabian1989}. In addition to the energy shifts, reflected photons take a longer path to reach us than those that travelled to the observer directly from the corona. This means that any fluctuations in the luminosity of the corona are first seen in the power-law continuum and are then, after a light-crossing delay, seen in the reflection spectrum. Measuring this \textit{reverberation lag} between direct and reflected radiation gives an extra diagnostic \cite{Uttley2014,Ingram2019}. In particular, measuring the lag and line profile together constrains the BH mass, since the reverberation lag depends on the actual geometric size of the disc and corona, whereas the line profile depends on the same sizes in units of gravitational radii \cite{Mastroserio2019}. Combining the two therefore calibrates the length of the gravitational radius and thus the mass.

The challenge of observing the reverberation lag is that we cannot completely isolate the direct and reflected signals. Any energy band we look in will include both direct and reflected photons. However, the direct and reflected signals have different spectra and so some energy bands will contain a greater fraction of reflected photons than others. We can therefore search for time lags between a reflection-dominated band and a continuum-dominated band. An extra complication, however, is thrown up by the time lags typically attributed to propagating accretion rate fluctuations discussed in Section \ref{sec:propfluc}. These lags are far larger than the expected reverberation lags, and they do not have the expected photon energy dependence. Whether or not they are specifically caused by propagating fluctuations, they are not due to light-crossing delays and are instead due to variability of the direct continuum spectral shape. For this reason, they are typically called \textit{continuum lags}. These continuum lags are so large at low Fourier frequencies that they prevent detection of reverberation lags. However, the magnitude of the continuum lags reduces with increasing frequency whereas the reverberation lag is roughly constant with frequency\footnote{The reverberation lag amplitude is approximately constant for frequencies well below the \textit{phase-wrapping} regime whereby the reverberation lag is comparable to the variability timescale itself ($1/\nu$). Above this frequency, the lag is oscillatory with frequency, analogous to car wheels appearing to rotate backwards or slowly forwards on a film due to beating between the rotation rate of the wheel and the frame rate of the camera that filmed it.}. This means that reverberation lags can be discovered at high Fourier frequencies.

\begin{figure}[b]
\sidecaption
\includegraphics[width=\textwidth,trim=1.5cm 0.0cm 1.5cm 0.5cm,clip=true]{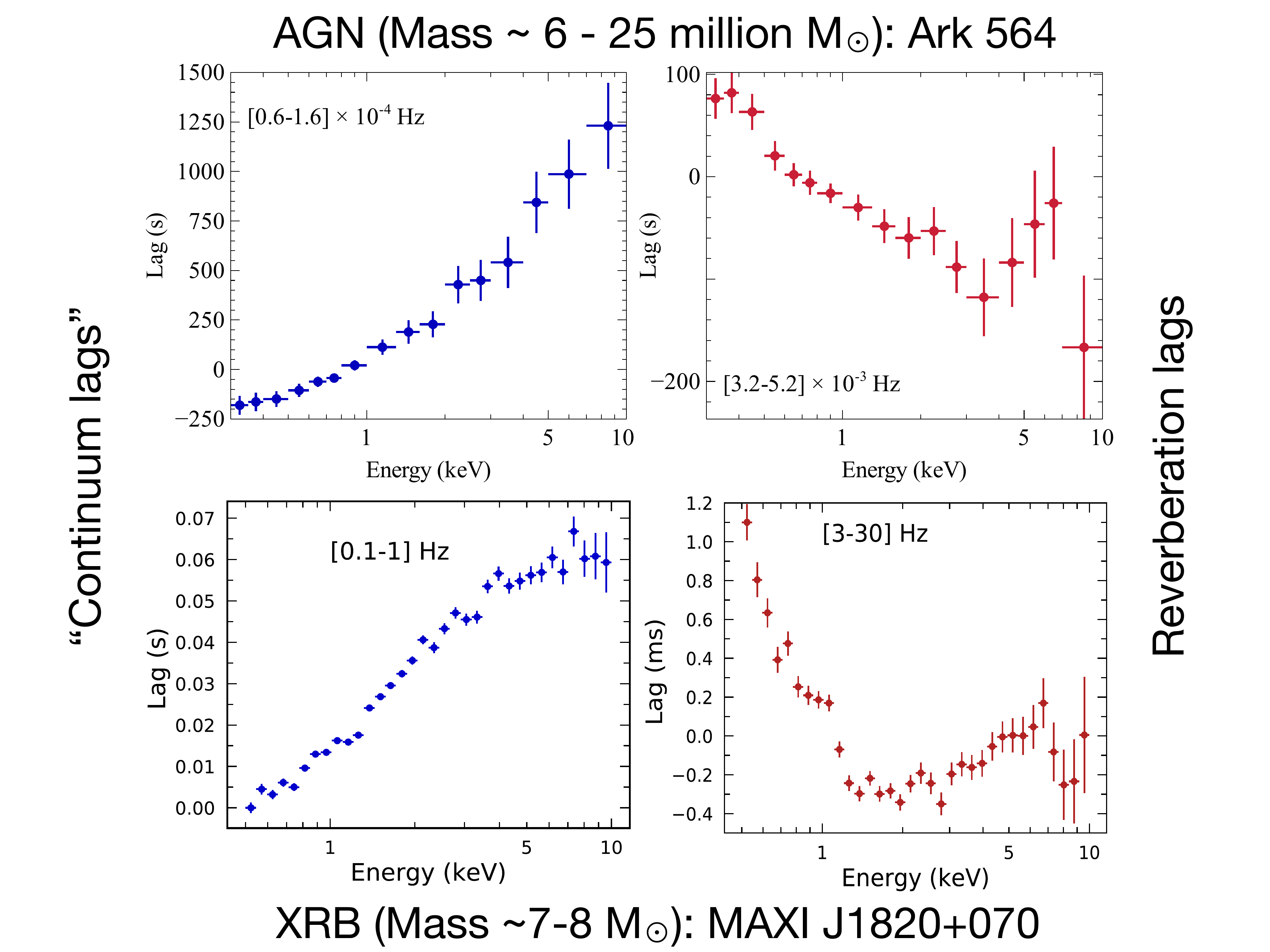}
\caption{Examples of observed lag vs energy spectra for an AGN (top row) and an XRB (bottom row) in low (left column) and high (right column) Fourier frequency ranges. At low frequencies, we see continuum lags and at high frequencies we see an iron line feature indicative of reverberation.}
\label{fig:revobs}       
\end{figure}

Fig \ref{fig:revobs} shows the observed time lag as a function of energy (the \textit{lag vs energy spectrum}) for an AGN (top row) and an XRB (bottom row), in a low Fourier frequency range (left column) and a high Fourier frequency range (right column). The lags are measured with respect to a common \textit{reference band}, such that photons with a larger lag value arrived after those with a smaller lag value. The choice of reference band has no effect on the shape of the lag vs energy spectrum, it simply defines the energy at which the lag is zero. For the low frequencies, we see that hard photons arrive after soft photons, following the $\propto \log(E_2/E_1)$ behaviour discussed in Section \ref{sec:propfluc}. For the high frequencies, we instead see an emission-like feature at $\sim 6.4$ keV. This is consistent with being due to reverberation: because of the iron line, the $\sim 6.4$ keV band includes a greater fraction of reflected photons than, say, the $2-3$ keV band. We see that the width of the line feature is similar for the XRB and the AGN, which is expected despite the AGN being $\sim 10^6$ times as massive because the energy shifts are independent of BH mass. The lag magnitude, however, is indeed around a million times larger for the AGN, since this depends linearly on mass. Note that the Fourier frequency ranges are also very different for the two objects, which is again expected since all physical timescales scale linearly with BH mass.

\begin{figure}[b]
\sidecaption
\includegraphics[width=\textwidth,trim=0.0cm 12.0cm 1.0cm 0.5cm,clip=true]{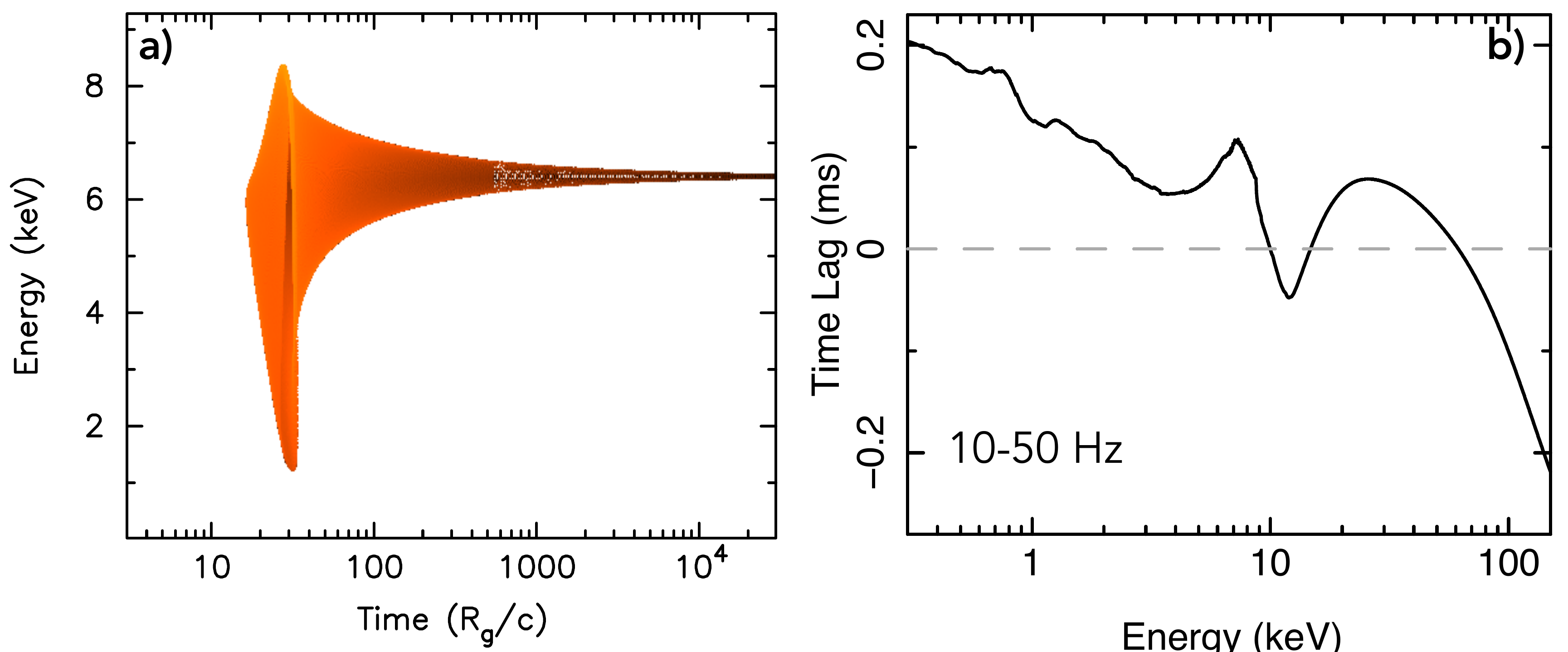}
\caption{\textit{a):} Impulse-response function for a rest frame reflection spectrum consisting only of a narrow iron line at 6.4 keV. The corona has been assumed to be a point-like source located on the BH spin axis some height $h$ above the BH (lamppost model). The source height is $h=6~r_g$, the disc is extends to the innermost stable circular orbit (ISCO), the BH dimensionless spin parameter is $a=0.9$ and the disc inclination angle is $i=70^\circ$. \textit{b):} Lag vs energy spectrum in the Fourier frequency range $10-50$ Hz calculated from the same parameters as in (a), with the BH mass additionally set to $M_{\rm bh}=10M_\odot$. The reference band is a narrow band centered on 10 keV, meaning that the lag trivially passes through zero at 10 keV (see grey dashed line).}
\label{fig:impresp}       
\end{figure}

Interpretation of the lag requires detailed modelling. As an illustration, consider that the amplitude of the line feature for the XRB (bottom right panel) is $\sim 0.4$ ms, but the actual reverberation lag between direct and reflected X-rays must be larger. This is because the iron line band still contains direct continuum photons and continuum-dominated bands such as $2-3$ keV still contain some reflection, which serves to dilute the lag between energy bands. The most computationally efficient way to calculate the lag vs energy spectrum from a model is to first calculate the \textit{impulse-response} function, $w(E,t)$. This captures the response of the reflection spectrum to a $\delta-$function flash in coronal luminosity, and Fig \ref{fig:impresp} shows an example. Here, the full reflection spectrum is replaced by a $\delta-$function iron line at $6.4$ keV for illustration, and the time axis is set such that the $\delta-$function flare in the direct continuum is observed at $t=0$. A short waiting time after the direct flare we see the first reflected photons, which are those that reflected from the inner edge of the disc. The line is therefore initially very broad due to Doppler shifts from rapid rotation at the inner disc edge and heavily reddened by strong gravitational redshift. We then see photons that reflected from progressively larger radii, and thus the iron line gets progressively narrower and dimmer.

In the simplest case, whereby the coronal luminosity varies but the spectral shape remains constant, we can write the time-dependent spectrum (specific energy flux) of the direct radiation as $d(E,t)=a(t)D(E)$. In this case, the time-dependent reflection spectrum can be written as a convolution between the normalisation $a(t)$ and the impulse-response. Taking the Fourier transform and applying the convolution theorem gives the following expression for the Fourier transform of the total time-dependent spectrum
\begin{equation}
    S(E,\nu) = A(\nu) \left[ D(E) + W(E,\nu) \right],
    \label{eqn:SEnu}
\end{equation}
where the Fourier transform of the impulse-response function, $W(E,\nu)$, is called the \textit{transfer function}. In order to calculate time lags, we must first define a reference band that lags are measured relative to. The reference band light curve, $f(t)$, is typically calculated by integrating the time-dependent flux over all available energy channels in order to maximise signal to noise \cite{Uttley2014,Ingram2019a}, but this is simply a design choice. The Fourier transform of the reference band light curve in this case is $F(\nu) = \int S(E,\nu) dE$. We can then take the \textit{cross-spectrum} between each subject band and the reference band, $C(E,\nu)=S(E,\nu)F^*(\nu)$. This is a complex quantity, and the phase lag between photons of energy $E$ and the reference band is given by its argument, $\phi(E,\nu) = {\rm arg}[C(E,\nu)]$. The time lag is simply related to the phase lag as $t_{\rm lag}(E,\nu)=\phi(E,\nu)/(2\pi \nu)$; although note the $2\pi$ phase ambiguity inherent to Fourier analysis. From Equation (\ref{eqn:SEnu}), the time lag can finally be written in terms of the transfer function \cite{Bambi2021}
\begin{equation}
    t_{\rm lag}(E,\nu) = \frac{1}{2\pi \nu} \arctan\left[ \frac{ {\rm Im}W(E,\nu) }{ D(E) + {\rm Re}W(E,\nu) } \right] - t_{\rm ref}(\nu),
\end{equation}
where $t_{\rm ref}(\nu)$ simply sets the zero point, ensuring that the time lag between the reference band light curve and itself is zero (i.e. changing the definition of the reference band \textit{only} changes the value of $t_{\rm ref}(\nu)$ and nothing else). Fig \ref{fig:impresp}b shows an example lag vs energy spectrum calculated using the model \textsc{reltrans} \cite{Ingram2019}.

The reflected component has a higher degree of linear polarization than the direct component, mainly due to the contribution of Compton scattered photons \cite{Matt1993b}. Other processes that contribute to the emergent spectrum (i.e. absorption and fluorescence) significantly complicate the picture. For instance, fluorescence lines are intrinsically unpolarized, resulting in a dip in polarization degree at emission line energies. The most advanced treatments of the polarization of the reflection signal in the literature \cite{Matt1993b,Dovciak2011} account for Compton scattering and photoelectric absorption in a slab of neutral material (H and He fully ionized, all heavier elements entirely in the form of bound atoms). The radiative transfer equation is solved by a Monte Carlo simulation that follows photons until they are either absorbed or escape the slab. Iron K$_\alpha$ and K$_\beta$ fluorescence lines are included, but fluorescence photons are assumed to immediately escape the slab. Since these calculations are nearly 30 years old, they are fairly primitive compared to the most advanced treatments without polarization information \cite{Ross2005,Garcia2010,Garcia2013}. For instance, the model \textsc{xillver} self-consistently calculates the ionisation state and temperature of the disc atmosphere, accounting for bound-bound, bound-free and free-free (including bremsstrahlung) processes, utilising the \textsc{xstar} atomic database \cite{Kallman2001}. The resulting spectrum therefore includes a superposition of self-consistently calculated absorption edges and emission lines from different ionic species of different elements. Compton scattering of fluorescence photons before escaping the slab is also taken into account, which will result in fluorescence lines having a low but now non-zero polarization. However, a version of \textsc{xillver} that includes polarization is still under development. 

Nevertheless, reflected X-rays are still expected to be more highly polarized than the directly observed X-rays, and we therefore expect reverberation lags to be present between high and low polarization populations of photons. The dawn of X-ray polarimetry-timing therefore provides an extra dimension to X-ray reverberation mapping, even if the signal may be too small for IXPE detect, requiring next generation instruments such as eXTP or beyond. Perhaps the neatest way to include polarization into the current reverberation formalism is to additionally calculate transfer functions for the Stokes parameters $Q$ and $U$. Let us represent the emergent reflection spectrum (i.e. Stokes $I$) in the disc restframe as $\mathcal{R}(E)$. The Stokes $Q$ and $U$ of the emergent reflection spectrum, again in the disc restframe, are then
\begin{eqnarray}
    \mathcal{Q}(E) &=& \mathcal{R}(E) p_r(E) \cos[ 2\psi_r(E) ] \nonumber \\
    \mathcal{U}(E) &=& \mathcal{R}(E) p_r(E) \sin[ 2\psi_r(E) ],
\end{eqnarray}
where $p_r(E)$ and $\psi_r(E)$ are respectively the energy-dependent (linear) polarization degree and angle of the reflected component. The (Stokes $I$) transfer function is given by \cite{Ingram2019}
\begin{equation}
    W(E,\nu) = \int_{\alpha,\beta} \epsilon(r) g^3(r,\phi) {\rm e}^{i 2\pi \tau(r,\phi) \nu} \mathcal{R}[E/g(r,\phi)] d\alpha d\beta,
\end{equation}
where the emissivity profile $\epsilon(r)$ specifies the radial dependence of irradiating flux, $g(r,\phi)$ is the energy shift experienced by photons as they propagate from disc coordinate $r,\phi$ to the observer ($g>1$ is a blue shift and $g<1$ is a red shift), and $\tau(r,\phi)$ is the time delay experienced by photons that reach the observer via reflection off the disc coordinate $r,\phi$ relative to those that travel directly from the corona to the observer. Here, $\alpha$ and $\beta$ are the impact parameters at infinity, and they can be interpreted as the horizontal and vertical coordinates on the image plane. A given position on the image plane (coordinates $\alpha$,$\beta$) can be mapped onto disc coordinates $r,\phi$ by tracing rays backwards in the Kerr metric from the image plane to the disc plane.

The above can be generalised by introducing the vector $\mathbf{W}(E,\nu)$ with its three components consisting of the transfer function for the three Stokes parameters. In addition to the energy shifts and time lags, we must account for the rotation of the polarization angle experienced by rays propagating from the disc to the observer (i.e. gravitational Faraday rotation). Rays that have polarization angle $\psi_r$ when they are emitted from disc coordinate $r,\phi$ will have polarization angle $\psi_r+\Delta\psi_r(r,\phi)$ by the time they reach the observer. This rotation can be calculated in the Kerr metric \cite{Connors1980,Dovciak2004}, and is accounted for by the following
\begin{equation}
    \mathbf{W}(E,\nu) =
    \begin{pmatrix}
    W_I(E,\nu) \\
    W_Q(E,\nu) \\
    W_U(E,\nu)
    \end{pmatrix}
    = \int_{\alpha,\beta} \epsilon(r) g^3 {\rm e}^{i 2\pi \tau \nu} ~\mathbf{T} ~\mathbf{S}_\mathcal{R}(E/g) ~d\alpha d\beta,
\end{equation}
where
\begin{equation}
\mathbf{S}_\mathcal{R}(E) = 
\begin{pmatrix}
\mathcal{R}(E) \\
\mathcal{Q}(E) \\
\mathcal{U}(E)
\end{pmatrix};~~~~~~~~
\mathbf{T} = 
\begin{pmatrix}
1 & 0 & 0 \\
0 & \cos[2\Delta\psi_r] & -\sin[2\Delta\psi_r] \\
0 & \sin[2\Delta\psi_r] & \cos[2\Delta\psi_r]
\end{pmatrix},
\end{equation}
and $\mathbf{T} ~\mathbf{S}_\mathcal{R}$ is a matrix multiplication. Note that the $r,\phi$ dependencies of $g$, $\tau$ and $\Delta\psi_r$ have been left implicit for brevity. The first component of this vector $W_I(E,\nu)=W(E,\nu)$ is simply the transfer function as introduced previously, and the other two entries are Stokes $Q$ and $U$ accounting for the rotation of the polarization angle in strong gravity.

We can then calculate two extra lag vs energy spectra: one between Stokes $Q$ and the reference band flux, and the other between Stokes $U$ and the reference band flux. These three lag vs energy spectra are related to the transfer function vector as
\begin{equation}
    \mathbf{t_{\rm lag}}(E,\nu) =
    \begin{pmatrix}
    t_{\rm lag, I}(E,\nu) \\ t_{\rm lag, Q}(E,\nu) \\ t_{\rm lag, U}(E,\nu)
    \end{pmatrix}
    = \frac{1}{2\pi \nu} \arctan\left[ \frac{ {\rm Im}\mathbf{W}(E,\nu) } { \mathbf{D}(E) + {\rm Re}\mathbf{W}(E,\nu) } \right] - \mathbf{t}_{\rm ref}(\nu),
    \label{eqn:tlag}
\end{equation}
where
\begin{equation}
    \mathbf{D}(E) = D(E)
    \begin{pmatrix}
    1 \\ p_D(E) \cos[2\psi_D(E)] \\ p_D(E) \sin[2\psi_D(E)]
    \end{pmatrix},
\end{equation}
and $p_D(E)$ and $\psi_D(E)$ are respectively the polarization degree and angle of the directly observed coronal emission. Ultimately, fitting for these three lag vs energy spectra instead of only one will provide tighter constraints and break model degeneracies. In particular, the rotations $\Delta\psi_r(r,\phi)$ can in principle be measured, which are of course completely inaccessible without polarization information, and are partially degenerate with the restframe polarization properties if only time-averaged polarimatric observations are considered. This in turn can provide better and more reliable constraints on BH mass and spin, and even constrain deviations from the Kerr metric to test GR itself \cite{Bambi2021a}.

\subsection{Quasi periodic oscillations}
\label{sec:QPOs}


BH and NS XRBs routinely display quasi-periodic oscillations (QPOs) in their X-ray flux \cite{Ingram2019b}. These features are best observed as a series of narrow, harmonically related peaks in the power spectrum of the X-ray flux. A pulsation in the flux would appear as a series of harmonically related $\delta-$function peaks in the power spectrum, whereas a QPO has some finite width. QPOs provide potentially very valuable diagnostics for the accretion flow and the compact object itself, since they are strong signals picking out some characteristic frequency of the system. However, their interpretation has proved challenging despite decades of interest. QPOs can be split into low frequency (LF) and high frequency (HF) QPOs. LF QPOs (frequency $\nu_{\rm qpo} \sim 0.1-20$ Hz) are often very strong signals, and thus are the best constrained observationaly. HF QPOs ($\nu_{\rm qpo} \gtrsim 60$ Hz) are typically weaker and less well observed, but have been the subject of great theoretical interest since their frequencies are commensurate with the orbital frequency in the vicinity of the compact object. HF QPOs from BH systems are very rare, but their apparent analogies in NS systems, kHz QPOs, are much more commonly observed. For both system classes, HF (kHz) QPOs are sub-divided into upper and lower HF (kHz) QPOs. It is common to observe a pair of kHz QPOs in the same power spectrum, in which case it is trivially clear that the upper kHz QPO is the one with the higher frequency. Doublets are much more rare for HF QPOs, but it is typically still possible to classify a single HF QPO as either lower or upper \cite{Ingram2014}.

\begin{figure}[b]
\sidecaption
\includegraphics[width=\textwidth,trim=0.0cm 0.5cm 0.0cm 12.0cm,clip=true]{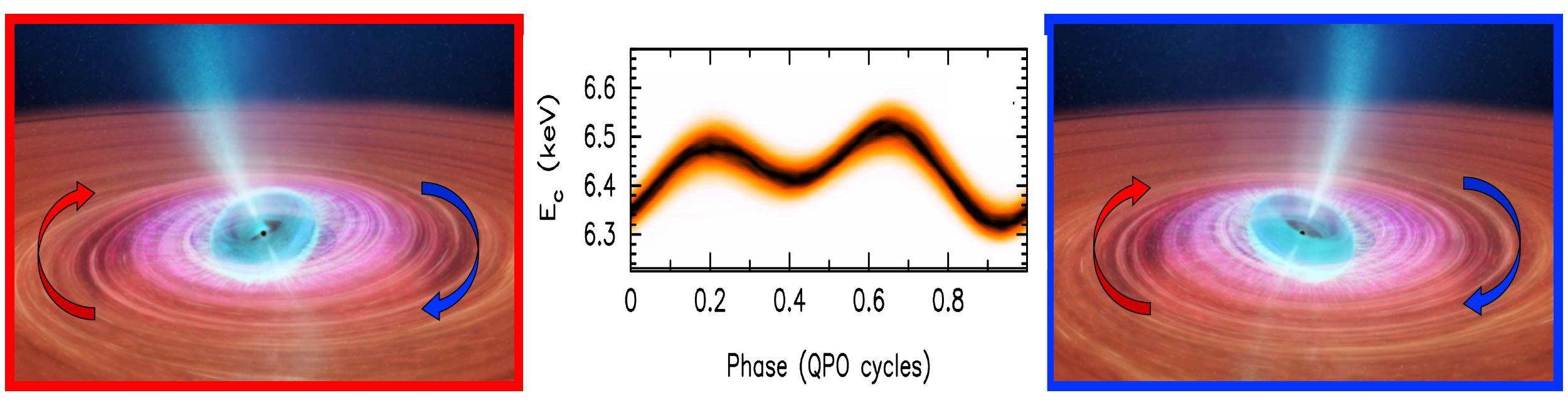}
\caption{\textit{Left and right:} Illustration of the precession model. The disc remains stationary while the corona (and the jet base) precess around the BH spin axis. \textit{Centre:} The iron line centroid energy of H 1743-322 as a function of QPO phase \cite{Ingram2017}. Colour represents posterior probability density (black is most probable). This centroid energy modulation is consistent with being due to the corona illuminating different disc azimuths at different phases of its precession cycle.}
\label{fig:prec}       
\end{figure}

Many LF QPO models have been suggested in the literature (see \cite{Ingram2019b} for a review). For instance, they could be due to Lense-Thirring precession of the corona \cite{Ingram2009}. This is a nodal precession, such that the orientation of the corona precesses around the BH spin axis due to the general relativistic frame dragging effect, and is illustrated schematically in Fig \ref{fig:prec}. Such precession requires the corona to be misaligned with the BH spin axis, and so the model requires there to be a modest misalignment between the binary orbital plane and the BH equatorial plane. The Lense-Thirring precession frequency of a test mass increases with proximity to the BH, but the dynamics of the accretion flow are additionally influenced by how warps are communicated throughout the flow. Analytic work and numerical simulations show that a thin disc forms a stationary configuration, with the inner disc aligned with the BH equatorial plane and the outer disc aligned with the binary plane \cite{Bardeen1975,Liska2019}. A low density accretion flow, such as is suggested as the origin of the corona, is instead able to precess at one average precession frequency \cite{Fragile2007,Liska2018}. The precession model is consistent with a number of observational properties of LF QPOs. For instance, we know that the coronal emission is modulated much more strongly than the disc emission \cite{Sobolewska2006}, and population studies indicate that the QPO properties depend on the inclination angle of the system \cite{Motta2015,vandeneijnden2017}. The strongest evidence in favour of the precession model is the observation shown in Fig \ref{fig:prec} (centre) of a QPO phase-dependence of the iron line centroid energy of H 1743-322 \cite{Ingram2016,Ingram2017}. This was a prediction of the precession model, since as the corona precesses it should shine predominantly on different disc azimuths. When the approaching/receding side of the disc is preferentially illuminated, the line will be bluer/redder due to Doppler shifts. However, other models can potentially explain this. For instance the accretion ejection instability (AEI) consists of spiral density waves rotating about the disc surface. The observed Doppler shifts to the iron line could therefore be due to the raised reflection emissivity from the spiral arms, rather than a precessing inner illuminator.

\begin{figure}[b]
\sidecaption
\includegraphics[width=\textwidth,trim=0.0cm 5.0cm 0.0cm 0.5cm,clip=true]{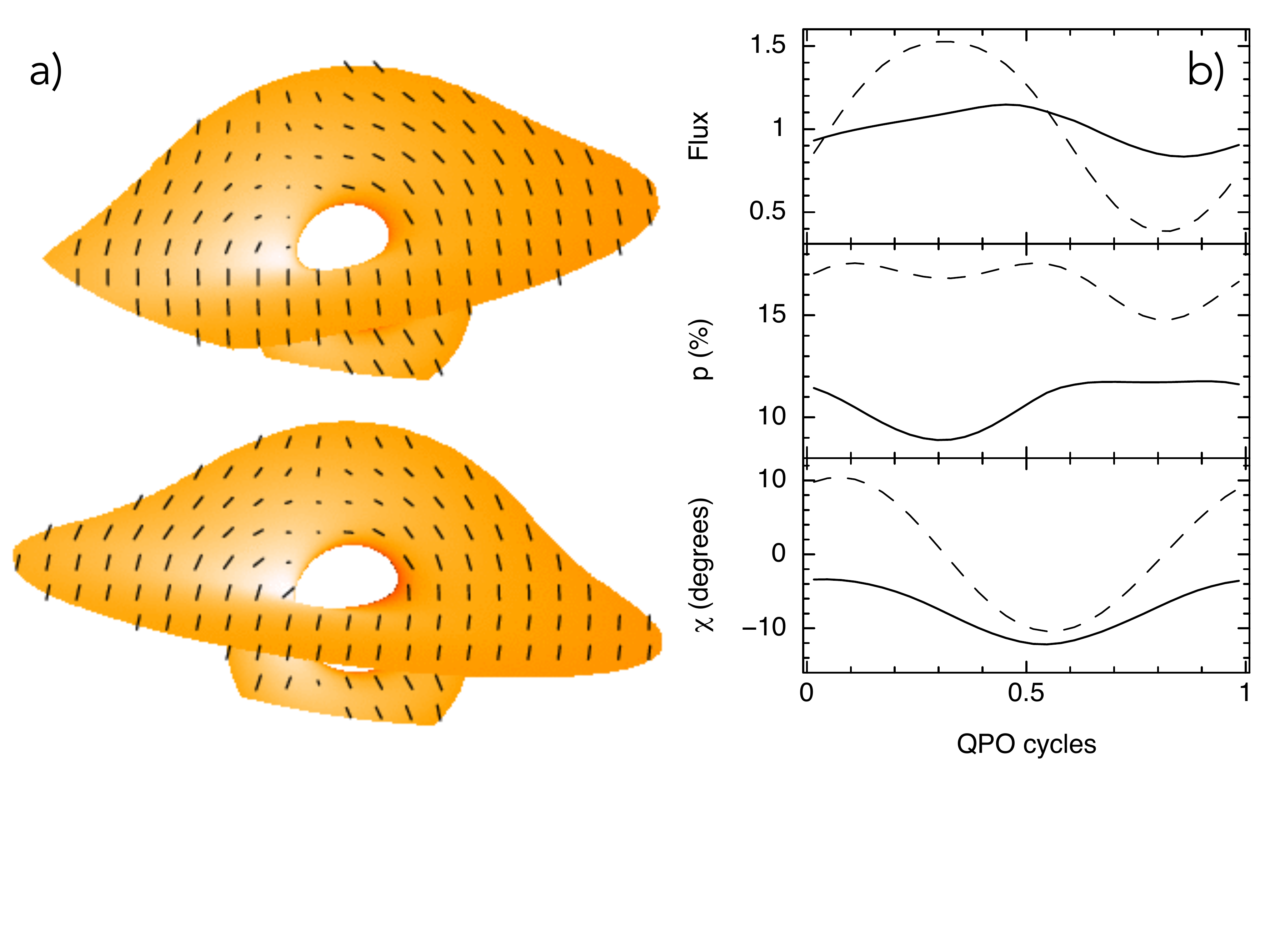}
\caption{\textit{a):} Ray-traced images of a precessing corona (modelled as a torus) at two precession phases. Colour indicates observed flux and the length and orientation of the black lines indicate polarization degree and angle of that pixel. The secondary image underneath the corona is due to rays from the bottom of the torus bending around the BH into our line of sight. A stationary accretion disc is assumed to be present that blocks some rays from the corona. \textit{b):} Precession phase dependence of the integrated quantities from the same calculation. Form top to bottom they are: flux, polarization degree and polarization angle. The inclination angle is $i=70^\circ$ and the outer radius of the corona is $20~r_g$. Adapted from \cite{Ingram2015a}.}
\label{fig:precpol}       
\end{figure}

X-ray polarimetry-timing will provide a `smoking gun' test of the precession model, since precession of the corona should cause swings in the X-ray polarization angle on the QPO period. The polarization degree is also expected to be modulated, since the polarization observed from a Comptonising slab depends on viewing angle. Fig \ref{fig:precpol} shows how the polarization properties of a precessing corona depend on precession phase \cite{Ingram2015a}. The corona is modelled as a torus that radiates in the same way as a Comptonising slab. Images at two QPO phases are in panel a. The colour represents observed X-ray flux and the length and orientation of the black lines represent respectively polarization degree and angle of that pixel. Panel b shows the total integrated flux, polarization degree and angle seen by the observer as a function of QPO phase. The dashed lines are for Newtonian gravity whereas the solid lines represent the calculation in full GR. As intuitively expected, the polarization angle swings back and forth as the orientation of the corona changes. The polarization angle modulation is washed out by light bending and other GR effects, but is still large enough that we may hope to make a detection. This figure is for an inclination angle of $70^\circ$. The size of the polarization swings is larger for lower inclinations. This is simple geometry: a precessing vector traces out a cone when viewed from the side, but rotates through $360^\circ$ when viewed form the top. The opposite is true for the flux, and it is indeed observed that stronger QPOs are observed from higher inclination sources \cite{Motta2015}. We may therefore naively think that we should target low inclination sources with weaker QPOs in their flux in order to try and detect the largest polarization modulations. However, the signal to noise with which we can measure polarisation increases with polarization degree, which itself is expected to increase with inclination angle. These two considerations trade off such that high and low inclinations are likely roughly equally good targets.

\begin{figure}[b]
\sidecaption
\includegraphics[width=\textwidth,trim=3.0cm 7.0cm 3.0cm 3.0cm,clip=true]{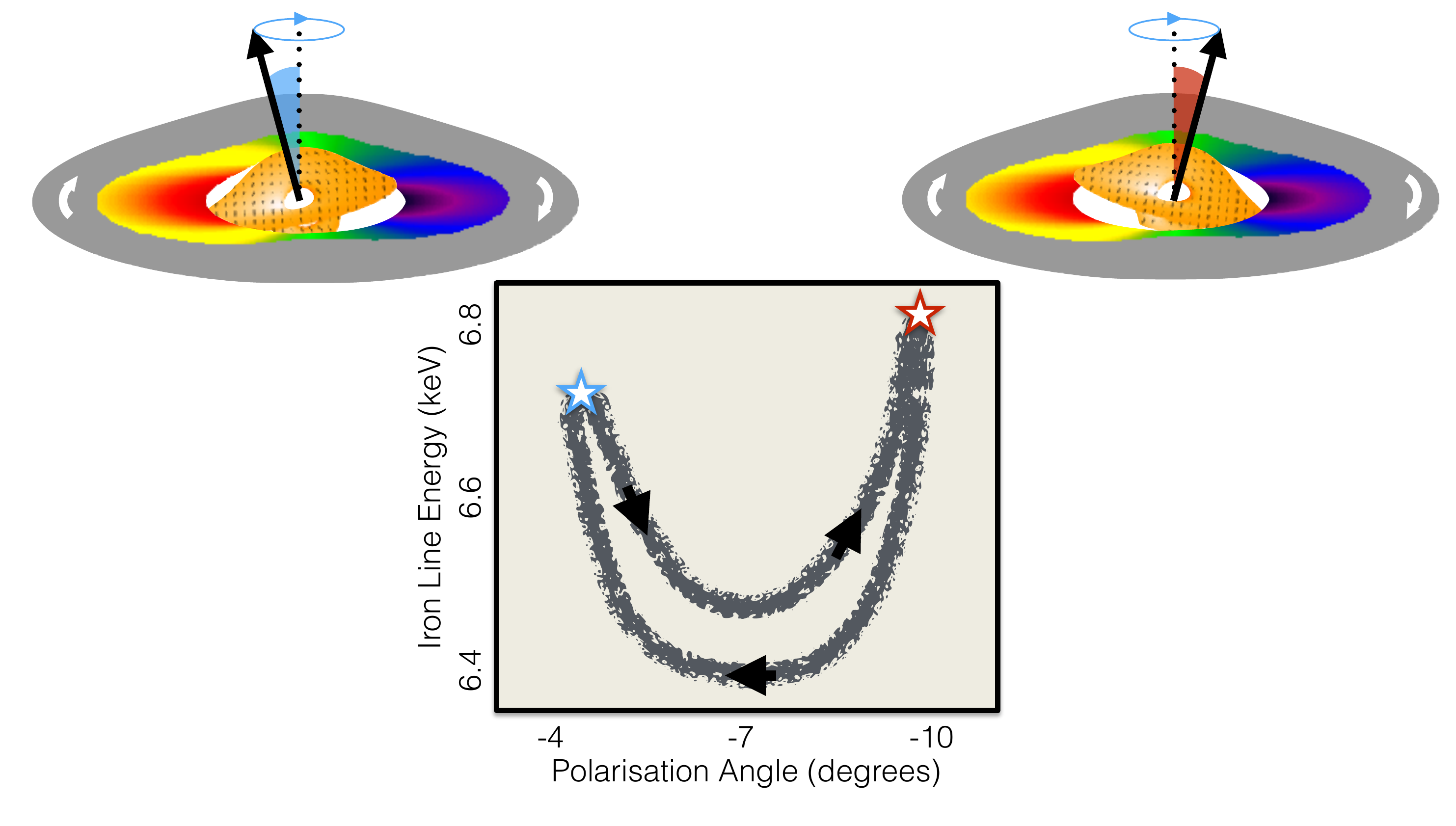}
\caption{Illustration of how the polarization angle and iron line centroid energy may co-evolve with QPO phase in the precession model. The two maxima in line energy per QPO cycle occur when the left and right sides of the disc are illuminated (top two images). The polarization angle associated each maximum is very different in the example shown: for one the corona is tilted to the left (blue) and for the other the corona is tilted to the right (red). This configuration would show up on a diagram of polarization angle vs line energy (bottom) as a smile shaped line.}
\label{fig:smile}       
\end{figure}

If it is possible to detect a modulation of polarization angle on the QPO period, a new diagnostic into the nature of the corona will be opened up. It will be particularly powerful to simultaneously track the QPO phase evolution of the polarization angle and the iron line centroid energy. Fig \ref{fig:smile} is a cartoon of what we might expect to see. In H 1743-322, we see two maxima in line energy per QPO cycle. One bright patch on the disc rotating about the surface once per QPO cycle would instead cause a line energy modulation with only one maximum per QPO cycle. There must therefore be two bright patches rotating about the disc surface. In this case, the bluest iron line occurs when the left and right (receding and approaching) sides of the disc are both illuminated (the top two images in Fig \ref{fig:smile}). This is because the blue shifted line emission from the approaching side is Doppler boosted and therefore dominates over the red shifted emission from the receding side \cite{Ingram2017}. The reddest line occurs when the front and back of the disc are illuminated, meaning that the observed line consists of only the core with no blue wing. Both configurations are present twice per QPO cycle, resulting in two minima and two maxima per cycle. If the variable illumination really is driven by precesison of the corona, then for the configuration illustrated in Fig \ref{fig:smile} the two maxima in line energy will be associated with very different polarization angles: for one maximum the corona is tilted to the left (maximum \ least negative polarization angle) and for the other it is tilted to the right (minimum / most negative polarization angle). A plot of polarization angle versus iron line energy will tell us how the corona illuminates the disc. For the example shown, the top and bottom of the corona are the brightest, which would lead to a smile shape being traced out in the polarization angle vs line energy diagram. If instead the sides of the corona are brighter than the top and bottom, we would expect a frown to be traced out on this diagram. The `smile/frown' diagram therefore has the potential to diagnose the angular emissivity profile of the corona.

The origin of HF QPOs is similarly uncertain. Perhaps the simplest interpretation is in the relativistic precession model \cite{Stella1998}, whereby a hotspot is orbiting in a slightly eccentric orbit at or near the inner radius of the accretion disc. In this model, the upper HF (or kHz) QPO is at the orbital period of the hotspot, and the lower HF (or kHz) QPO is at the periastron precession period. Perhaps the greatest success of this model comes from the one observation in the entire RXTE archive that features the full triplet of LF plus both HF QPOs in the same power spectrum \cite{Motta2014}. Applying the relativistic precession model to this observation yields a precise BH mass measurement ($M_{\rm bh}=5.31\pm0.07 M_\odot$) that agrees remarkably well with the dynamically measured value ($M_{\rm bh}=5.40\pm0.3 M_\odot$). If the HF QPOs really are driven by hotspots orbiting the BH (or NS for the case of kHz QPOs), then we expect the polarization degree and angle to be modulated on the upper HF QPO period \cite{Beheshtipour2016}. This is because gravitational Faraday rotation (i.e. $\Delta\psi_r(r,\phi)$ from Section \ref{sec:reverb}) depends strongly on azimuth for a spinning BH (due to the frame dragging effect), so a hotspot with constant emission properties will appear to change its polarization angle as it moves from one side of the BH to the other. Such modulations will be very challenging to detect with IXPE, but detection may be possible with eXTP \cite{DeRosa2019}.

\subsection{Blazars}
\label{sec:blazars}

Blazars are AGN viewed down the barrel of the jet \cite{Hovatta2019}. Since the electrons in the jet are therefore travelling towards us at relativistic velocities, their radiation is strongly beamed. Blazars are therefore observed to be very bright and jet-dominated. Since the jet is thought to be collimated by a twisted magnetic field, observations of polarization variability provide insights into the importance of magnetic turbulence to the jet dynamics. Observations of optical linear polarization reveal high variability in polarization degree and angle on timescales as short as sub-day \cite{Marscher2021}. The polarization degree on a given day is typically in the range $\sim 1-30\%$, but averaging over longer timescales gives a more modest polarization ($p\sim 4-10\%$) due to cancelling of misaligned polarization vectors from different days. The average polarization angle tends to roughly align with the spatially resolved  jet orientation. In some objects, the polarization angle only varies in the range $\pm 20^\circ$ around the mean, whereas in others large $>180^\circ$ swings are observed. This indicates a picture whereby significant ordered and turbulent components to the magnetic field are both present in the jet.

\begin{figure}[b]
\sidecaption
\includegraphics[width=\textwidth,trim=30.0cm 5.0cm 16.0cm 8.0cm,clip=true]{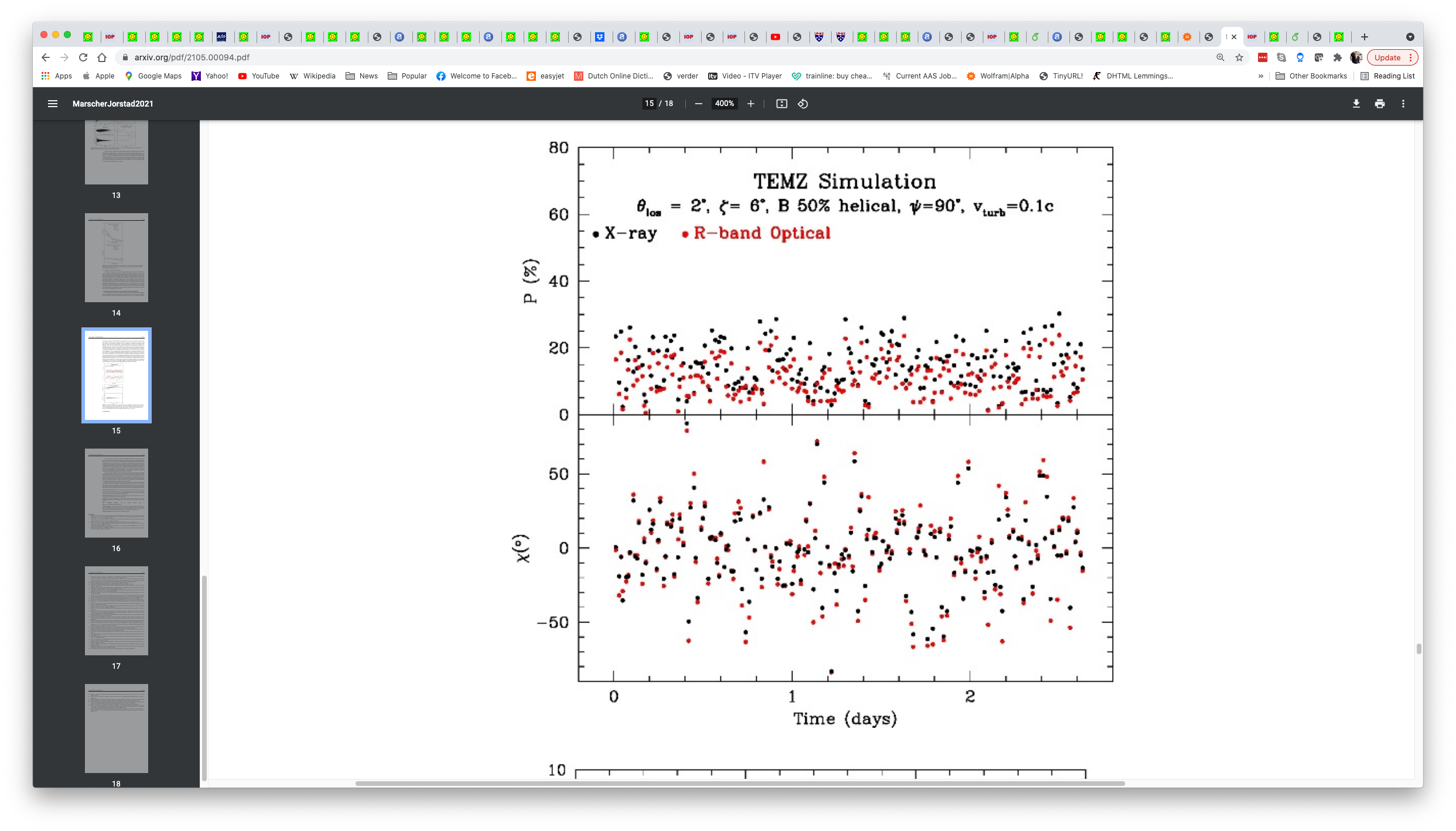}
\caption{Optical (red) and X-ray (black) polarization degree (top) and angle (bottom) as a function of time predicted for the blazar Mrk 421 by the TEMZ code. The code, described in more detail in the text, simulates a jet with both a stable helical and a turbulent magnetic field component, plus a conical re-collimation shock. Model parameters are: inclination angle: $\theta_{\rm los}=2^\circ$, half opening angle of the conical shock: $\zeta=6^\circ$, stable and turbulent components of the magnetic field have equal strength, pitch angle of the helical field: $\psi=90^\circ$, and maximum turbulent velocity: $v_{\rm turb}=0.1 c$. Reproduced from \cite{Marscher2021}.}
\label{fig:blazar}       
\end{figure}

The X-ray polarization is also predicted to be highly variable. Fig \ref{fig:blazar} shows the results of simulations with the turbulent extreme multi-zone (TEMZ) model \cite{Marscher2017}. This model simulates electrons with some bulk Lorentz factor being accelerated up a cylindrical section of jet. The magnetic field has a stable helical component and a turbulent component, and a conical re-collimation shock front with half opening angle $\zeta$ is included. The parameters in the figure were chosen to roughly reproduce the optical polarization properties of Mrk 421, thereby providing a prediction for the X-ray polarization properties of the same source, which will be an IXPE target. The pitch angle of the helical magnetic field is set to $90^\circ$ (this is $\psi$ in the figure legend, not to be confused with polarization angle for which this symbol is used elsewhere), indicating a maximally tight spiral that leads to the polarization angle aligning on average with the jet orientation. We see that the polarization degree is slightly higher in the X-rays than in the optical, and the X-ray polarization properties are similarly variable to their optical counterparts. Making observations of such variable X-ray polarization, and comparing with the optical signal, will therefore provide a new way to constrain jet models.

\section{Observational techniques}

In this Section, we will discuss how to detect and analyse rapid variability in X-ray polarization degree and angle. As a preface, it is worth first considering how X-ray polarization is measured. This topic is covered in far more detail elsewhere in this book, but in the interest of being self-contained let us go over the basics here. X-ray polarimeters detect individual photons. For each photon, the \textit{modulation angle}, $\tilde{\psi}$, is measured. For photoelectric polarimeters such as IXPE, each detected photon excites a photo electron in the detector. The modulation angle is measured from the initial direction of this photo electron. If the source has a polarization degree $p$ and angle $\psi$, the probability density of measuring a modulation angle $\tilde{\psi}$ is given by the \textit{modulation function}
\begin{equation}
    f(\tilde{\psi}) = \frac{1}{2\pi} \left\{ 1 + \mu~p~\cos\left[ 2(\psi-\tilde{\psi})\right] \right\},
    \label{eqn:modfunc}
\end{equation}
where $\mu$ is the modulation angle, which is a property of the polarimeter that describes its response to a $100\%$ polarized signal. An X-ray polarimeter has $\mu>0$, which enables two ways of measuring polarization. One can either make a histogram of counts vs modulation angle and fit the above function to it. The peak of the histogram occurs at $\psi$ and the amplitude is related to $p$. Note that the histogram always has two peaks. This is because polarization is only defined on an interval of $\pi$ radians, which is because `upwards' and `downwards' polarization are indistinguishable from one another; they are both simply examples of vertical polarization. The other way is to calculate Stokes parameters. If we detect $N$ photons such that the $k^{\rm th}$ photon to be detected has an associated modulation angle $\tilde{\psi}_k$, then Stokes $Q$ and $U$ are \cite{Kislat2015}
\begin{equation}
    Q = \frac{2}{\mu} \sum_{k=1}^N \cos( 2 \tilde{\psi}_k ); ~~~~~~ U = \frac{2}{\mu} \sum_{k=1}^N \sin( 2 \tilde{\psi}_k ),
    \label{eqn:QUobs}
\end{equation}
and the polarization degree and angle can be calculated from the usual relations
\begin{equation}
    p = \frac{\sqrt{Q^2+U^2}}{N}; ~~~~~~ \psi = \frac{1}{2} \arctan\left( \frac{U}{Q} \right).
\end{equation}
After this crash-course in detecting polarization, we can move onto detecting \textit{variability} in polarization.


\subsection{Direct measurement}

The most obvious way to study variability in polarization properties is to measure the polarization degree and angle for discrete intervals in time and build up light curves for both. However, many photons are required in order to detect polarization with high statistical confidence. For a given count rate, there is thus a minimum exposure time in which the required number of photons can be collected by our detector. This therefore sets a minimum time bin interval for a light curve of polarization degree or angle. We can calculate this minimum time interval by considering the minimum detectable polarization (MDP). This is the minimum polarization degree for which it is possible to make a detection with some confidence $\mathcal{L}$. In other words, there is a probability $1-\mathcal{L}$ of measuring a polarization degree ${\rm MDP}(\mathcal{L})$ from an unpolarized source. The MDP is given by \cite{Weisskopf2010}
\begin{equation}
    {\rm MDP}(\mathcal{L}) = \frac{\sqrt{-4\ln(1-\mathcal{L})}}{\mu \langle s \rangle} \sqrt{\frac{\langle s \rangle + \langle b \rangle}{T}},
\end{equation}
where $\langle s \rangle$ and $\langle b \rangle$ are the mean source and background count rates respectively, and $T$ is the exposure time. It is typical to consider $\mathcal{L}=99\%$ confidence, for which ${\rm MDP}_{99\%} = 4.29 / ( \mu \langle s \rangle ) \sqrt{ (\langle s \rangle + \langle b \rangle) / T}$. We can re-arrange to obtain the minimum time interval required to make a $99\%$ confidence detection of a source with polarization degree $p$
\begin{equation}
    T_{99\%} = \left( \frac{4.29}{\mu~p} \right)^2 \frac{\langle s \rangle + \langle b \rangle}{ \langle s \rangle^2}.
\end{equation}

We can now use this equation to determine the variability timescales accessible to IXPE ($\mu\approx 0.3$) for a few different object classes. For XRBs, we may optimistically expect $p=10\%$, $\langle s \rangle = 100$ c/s, $\langle b \rangle = 0$, giving $T_{99\%}\approx 3.5$ minutes. Therefore, we will not be able to access the second to sub-second timescales of interest for XRBs with IXPE by simply directly making $p$ and $\psi$ light curves. For (Type 1) AGN, similarly optimistic numbers are $p=10\%$, $\langle s \rangle = 1.5$ c/s, $\langle b \rangle = 0$, giving $T_{99\%}\approx 3.8$ hours. Although this is still large, mass scaling means that variability timescales of interest may be just about accessible for the most massive AGN, providing that the most massive examples of the brightest AGN in the sky (e.g. IC 4329A) also happen to have a fairly high polarization degree. However, this estimate is likely too optimistic, since the brightest AGN are Type 1 and are therefore viewed at a reasonably low inclination angle, for which lower polarization is expected \cite{Chandrasekhar1960}. Still, AGN are closer to being accessible to direct methods than XRBs because their flux is only $\sim$2 orders of magnitude lower whereas their characteristic variability timescales scale linearly with BH mass, which is $\sim 5-8$ orders of magnitude higher. A sub-class of AGN for which it may well be possible to directly detect X-ray polarization variability with IXPE is blazars. As discussed in Section \ref{sec:blazars}, sub-day variability is observed in the optical polarization degree and angle of these objects, and similarly rapid variability is also expected in the X-ray band (see Fig \ref{fig:blazar}). In fact, for these objects shortening the time bin interval could make detection \textit{easier}. This is because the polarization angle can change so much that averaging over a long observation can significantly dilute the measured polarization degree.

Blazars aside, it is clear that for most science cases discussed in Section \ref{sec:theory}, a new method will be required other than simply directly constraining a light curve of $p$ and $\psi$.

\subsection{Stokes parameters}

Particularly astute readers may be wondering why we need statistically significant detections of polarization in each time bin in order to analyse light curves of polarization degree and angle. After all, we can make light curves of the count rate on arbitrarily short time scales and never need to worry that the source is not significantly detected in each of the separate time bins. The answer is simply down to the probability distribution of $p$ and $\psi$ versus that of the count rate. The count rate is Poisson distributed, the effects of which are thankfully rather benign. We can, for example, make a power spectrum of a time series of count rate recorded with arbitrarily short time bins, and all we need to do to correct for the Poisson nature of the detections is to subtract a constant Poisson noise level off the power spectrum (this is assuming no instrumental effects such as deadtime). The probability distributions of $p$ and $\psi$ are much less friendly. Still, one may reasonably think that we could navigate this problem if only we were clever enough. For instance, can we not extract a light curve for $p(t)$ with arbitrarily short time bins, then calculate the power spectrum of this light curve and finally correct for the effects of the probability distribution of $p(t)$, analogous to subtracting a constant Poisson noise component off the flux power spectrum? The problem seems difficult but not insurmountable until one appreciates that the polarization degree and angle are both completely undefined in time bins with no counts. There is surely no way to correct for such a catastrophic loss of information.

The Stokes parameters are a little more promising, since it is not necessarily a problem for a light curve of $Q$ or $U$ to have time bins with zero counts. The statistics for $Q$ and $U$ are still, however, complex. Imagine we extract observed $Q(t)$ and $U(t)$ using Equations (\ref{eqn:QUobs}). We can calculate the power spectra of these two time series by taking the modulus squared of the Fourier transform. In order to get Gaussian error bars, we must average the power spectrum. We can average over different light curve segments (\textit{ensemble averaging}) and over adjacent Fourier frequencies \cite{vanderKlis1989}, as is described in detail elsewhere in this book. Using angle brackets to denote this averaging process, we can write the two power spectra as $P_Q(\nu)=\langle |Q(\nu)|^2\rangle$ and $P_U(\nu)=\langle |U(\nu)|^2\rangle$. The challenge is to correct for the small number of counts in each time bin to convert from e.g. $P_Q(\nu)$ to an unbiased estimate of the true $|Q(\nu)|^2$ that would be measured if we had $N(t)\rightarrow\infty$ counts in each time bin. 

\begin{figure}[b]
\includegraphics[width=0.5\textwidth,trim=1.0cm 1.0cm 2.0cm 11.0cm,clip=true]{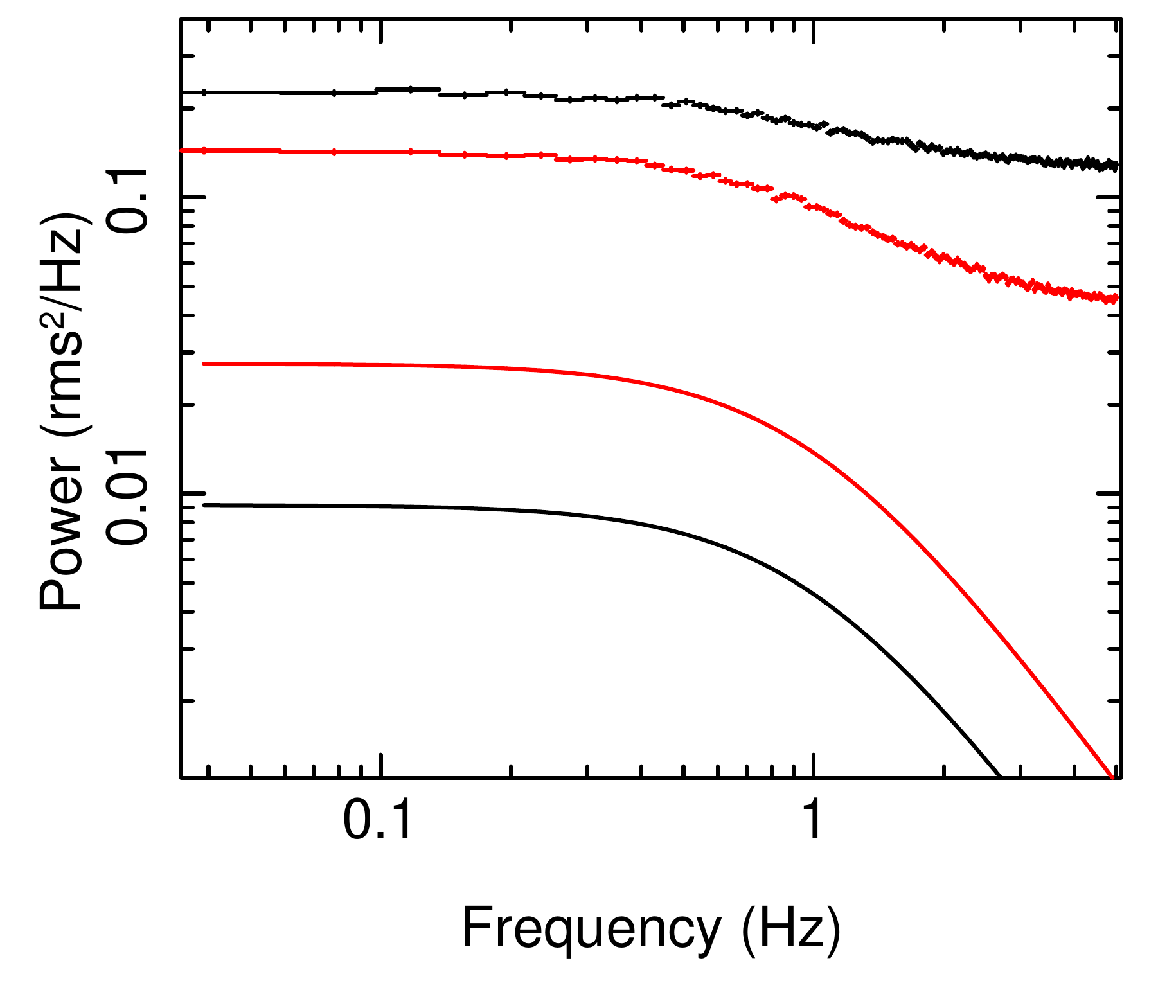}
\includegraphics[width=0.5\textwidth,trim=1.0cm 1.0cm 2.0cm 11.0cm,clip=true]{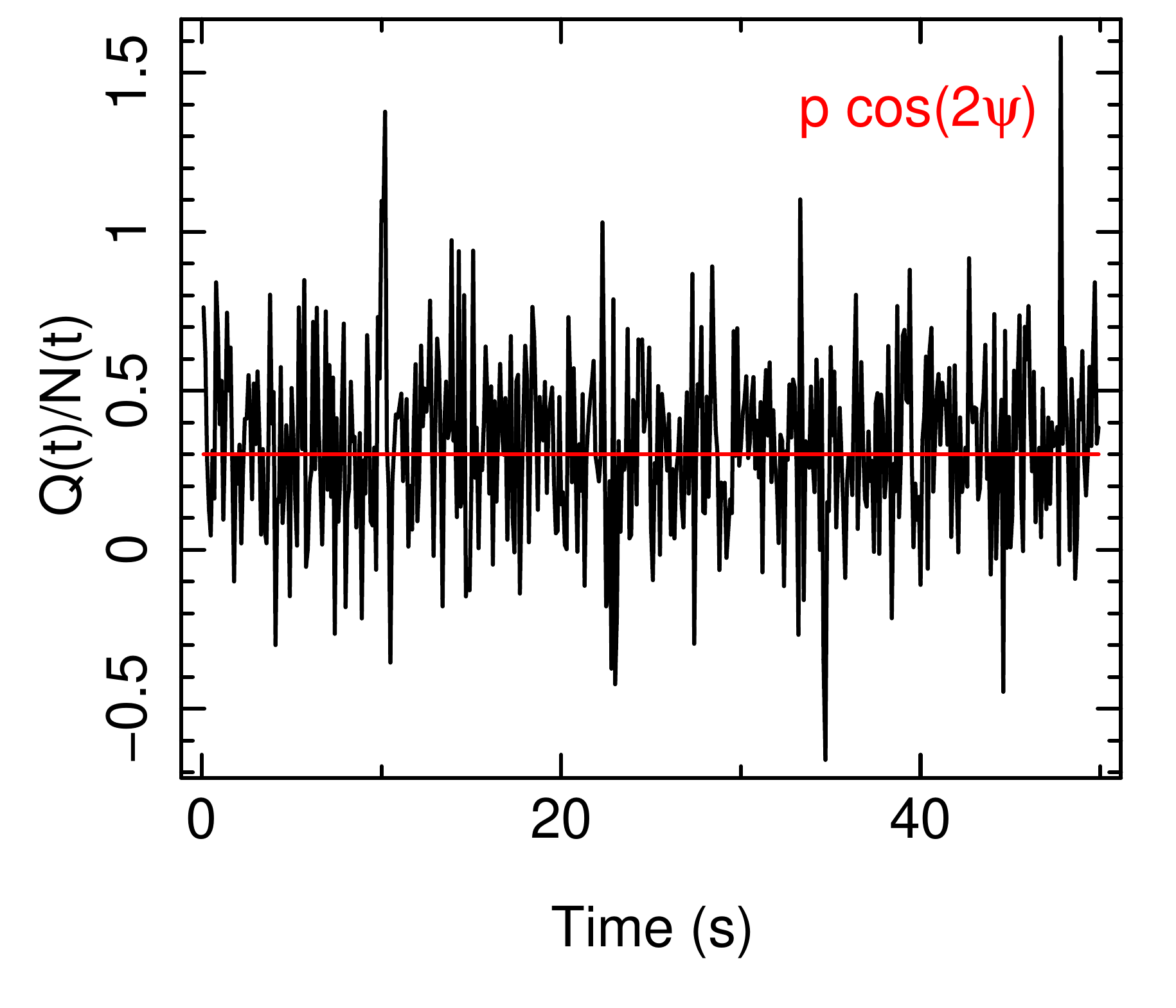}
\caption{Results of a simulation of the Stokes parameters, $Q(t)$ and $U(t)$, varying in time. \textit{Left:} Power spectrum of $Q(t)$ (black) and $U(t)$ (red) recovered from the simulation (points with error bars) and predicted analytically for a count rate approaching infinity (solid lines). Even though the simulation assumes a large count rate for a polarimeter ($1000$ c/s), constant $p$ and $\psi$, and very high polarization ($p=0.6$, $\mu=0.6$), Poisson statistics still have an enormous affect on the shape of these power spectra. \textit{Right:} Short segment of the simulated light curve for $Q(t)$, divided by the counts in each bin, $N(t)$. For a count rate approaching infinity, the simulation (black) would converge to the red line. We instead see a large ampitude of variability around the red line.}
\label{fig:Pq}       
\end{figure}

Fig \ref{fig:Pq} shows the results of simulations ran to explore the nature of this correction. A flux time series, $I(t)$, has been simulated using the maximum entropy method of Timmer \& Koenig \cite{Timmer1995}. This ensures that the flux time series has some target power spectrum (in this case, a zero centred Lorentzian with an integrated fractional root mean square amplitude of $0.4$ and a half width at half maximum of $1$ Hz), but the Fourier phases are all completely random. The simulated flux is then converted to count rate for $2^{20}$ time bins, each of $0.1$s duration, by selecting Poisson distributed random variables. For each photon in the time series, a random variable was selected for $\tilde{\psi}_k$ from the distribution defined by Equation (\ref{eqn:modfunc}), assuming constant polarization degree and angle. Stokes parameters were then calculated for each time bin from Equations (\ref{eqn:QUobs}). Extremely optimistic parameters are used in order to demonstrate that small number statistics would still be a problem even if we had an X-ray polarimeter better than even the next generation of instruments and a source far more polarized than expectation: $\langle s \rangle = 1000$ c/s, $p=60\%$, $\psi=30^\circ$, $\mu=0.6$. Fig \ref{fig:Pq}a shows $P_Q(\nu)$ (black points) and $P_U(\nu)$ (red points) recovered from the simulation, where the average has been over $2^{12}$ segments, each containing $2^8$ time bins. The lines (same colour scheme) show the expectation in the absence of Poisson noise; i.e. if there were so many photons that $\psi=30^\circ$ and $p=60\%$ could be recovered for each time bin of the simulated data with high precision. This would give $P_Q(\nu) = |I(\nu)|^2 p^2 \cos^2(2\psi)$ and $P_U(\nu) = |I(\nu)|^2 p^2 \sin^2(2\psi)$. The huge difference between the simulation and this idealised case demonstrates that counting statistics hugely increase the variability amplitude of the Stokes parameters. The leading order effect is that $\tilde{\psi}_k$ is selected from a very broad probability distribution (i.e. Equation \ref{eqn:modfunc}) and so $\sum_{k=1}^N \cos(2\tilde{\psi}_k)$ and $\sum_{k=1}^N \sin(2\tilde{\psi}_k)$ both take a very wide range of values unless $N$ is huge. This is demonstrated in Fig \ref{fig:Pq}b, where $Q(t)/N(t)$ is plotted against time. For a count rate approaching infinity, this would be constant in time, equal to $p \cos(2\psi)$ (red line). We see that for the simulation (black line), even for the relatively large count rate of $1000$ c/s, the vale of $Q(t)/N(t)$ instead varies greatly around an average value of $p \cos(2\psi)$. An analytic model for how this effect influences the power spectrum of the Stokes parameters, even in this simplified case discussed with constant polarization degree and angle, is currently elusive. We therefore must find other ways to study the timing properties of the polarization properties.

\subsection{Pulsations}

For periodic signals such as pulsations, the solution to the above problem is conceptually straight forward: the pulse phase is determined as a function of photon arrival time by folding on the pulse period and from that $I$, $Q$ and $U$ (and therefore $p$ and $\psi$) can be calculated for each of $K$ phase bins. Since photons are collected in each phase bin over the course of many, many pulse periods, a significant detection can be achieved in each pulse phase bin. The number of phase bins that the signal can be split into (if $99\%$ confidence detections are desired) is
\begin{equation}
    K = \left( \frac{\mu~p}{4.29} \right)^2 \frac{ \langle s \rangle^2}{\langle s \rangle + \langle b \rangle}~T,
\end{equation}
where $T$ is the total exposure time of the observation. Of course, in reality there are many complications to pulsar timing such as correcting for the orbital motion of the pulsar if it is in a binary system. These considerations are discussed in detail elsewhere in this book. The key point here, however, is that none of these considerations are new to polarimetry. Calculating the pulse phase-dependence of $p$ and $\psi$ is an almost trivial extension to the (non-trivial) process of calculating a pulse profile.

\subsection{Phase-folding of QPOs}

\begin{figure}[b]
\includegraphics[width=\textwidth,trim=0.0cm 1.0cm 0.0cm 2.0cm,clip=true]{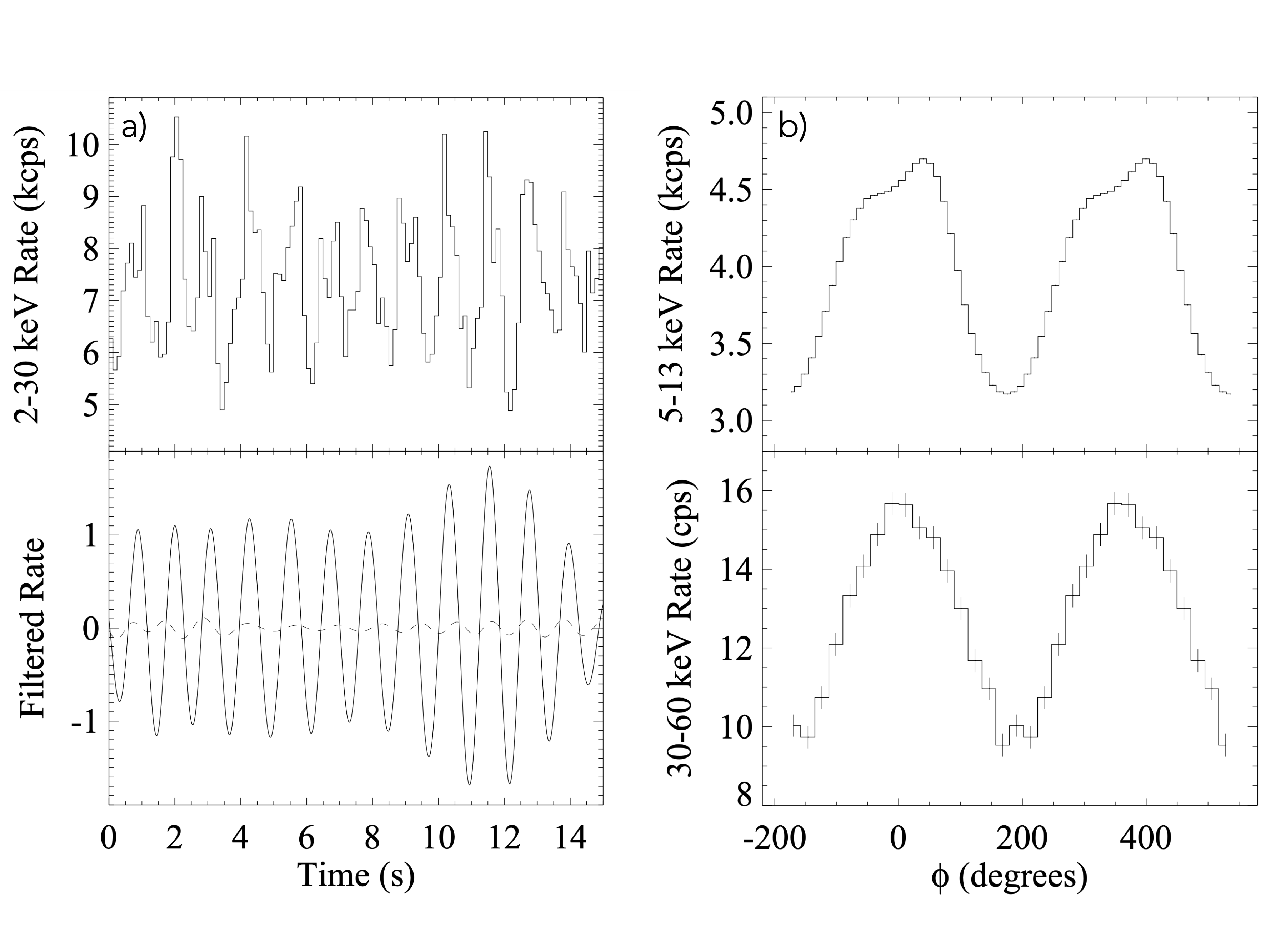}
\caption{\textit{Left:} Raw (top) and filtered (bottom; solid line) light curve of an RXTE observation of GRS 1915+105 exhibiting a strong LF QPO. The filtered light curve is an estimate for the QPO signal in the absence of the broad band noise. The dashed line is a simulated pure Poisson noise light curve that has been filtered in the same manner as the observed light curve. \textit{Right:} Phase-folded QPO profiles using PCA (top) and HEXTE (bottom) data. QPO phases are assigned from the filtered light curve. Adapted from \cite{Tomsick2001}}
\label{fig:filter}       
\end{figure}

Phase folding is much more difficult for QPOs, since the phase angle does not increase with time in a predictable way. Indeed, the QPO phase is observed to drift on a random walk away from that of a strictly periodic signal \cite{Morgan1997}. QPOs are also coincident with broad band noise, and so the challenge of QPO phase-folding is to first disentangle the QPO signal from the broad band noise signal and then use this estimated `pure' QPO signal to assign QPO phase values to every time bin \cite{Tomsick2001,vandeneijnden2016,Henric2016,DeRosa2019}. The first step is typically approached by filtering the raw light curve. Fig \ref{fig:filter}a shows the results of treating a light curve observed from GRS 1915+105 with a Kaiser filter, which discards all variability on frequencies $20\%$ above and below the QPO fundamental centroid frequency \cite{Tomsick2001}. We see that long term trends and rapid fluctuations in the raw light curve (top) are not present in the filtered light curve (bottom). Other choices of filter can be made. For example, a Wiener filter \cite{vandeneijnden2016}, which is optimal in the case whereby the signal that the user wishes to keep is completely uncorrelated with the signal (noise) that the user wishes to discard. This is not the case here, since the QPO and broadband noise \textit{are} correlated (e.g. \cite{Heil2011}), and so a Wiener filter is not formally optimal. It still, however, represents a sensible choice of filter.

The next step is to assign QPO phase to the filtered light curve. This can be done, for example, by looking for minima in the filtered QPO light curve. Fig \ref{fig:filter}b shows the folded light curves recovered from the filtered light curve shown in Fig \ref{fig:filter}a by defining a QPO phase of zero as all the times when the numerical derivative of the filtered light curve crosses zero and the numerical second derivative is negative (i.e. minima in the filtered light curve) \cite{Tomsick2001}. Here, QPO phases have been assigned to each time bin by approximating that the QPO period is constant between QPO minima. Perhaps a more sophisticated method is to take the Hilbert transform \cite{ Huang2008} $y(t)$ of the filtered light curve $x(t)$. In this case, the instantaneous QPO phase is $\phi(t) = \arctan[y(t)/x(t)]$ \cite{DeRosa2019}, meaning that the QPO period is no longer only estimated once per QPO cycle but instead it is effectively estimated anew for each time bin. Regardless of the method used, once QPO phase has been estimated for each time bin, it is simple to calculate the Stokes parameters for $K$ QPO phase bins and from that build up a QPO waveform of flux, polarization degree and polarization angle vs QPO phase.

Phase-folding methods have the advantage of being intuitive: we end up with a waveform for the polarization properties as a function of QPO phase with error bars that are simple to understand. However, there are many downsides. First of all, there are many design decisions. For example, what filter should we use and what are the effects of using the wrong one, and is it even possible to disentangle the QPO from the broadband noise with something as simple as a filter? The step of assigning QPO phases also brings design decisions, but we at least have a way to assess how well the process has worked because each mistake we make in assigning phase serves to reduce the amplitude of the folded light curve. Therefore, the closer the amplitude of the folded light curve is to the QPO amplitude measured from the power spectrum, the better the job we have done assigning phases. However, an extra complication is that phase-folding methods inevitably under predict the amplitude of higher harmonics. This is because the method estimates the phase of the fundamental component, but the second harmonic, and higher harmonics, are not perfectly coherent with the fundamental \cite{Ingram2015}. In other words, whereas a pulse profile stays constant in time, the QPO waveform varies in time around some well defined mean profile. Therefore, estimates of QPO phase that maximise the amplitude of the fundamental component in the folded light curve do not maximise the amplitude of the higher harmonics! Another downside is that phase-folding can only be used for strong QPOs. The example shown in Fig \ref{fig:filter} is of an extremely strong LF QPO that can be seen by eye in the light curve. Weaker LF QPOs will prove much harder to phase fold, and HF QPOs will be even more challenging still. The final downside is that phase-folding is of course limited to QPOs. What if we wish to conduct polarimetry-timing of the broad band noise; for instance polarized reverberation mapping or searching for time lags between highly polarized jet emission and weakly polarized corona emission? It turns out that there is a general method with well understood statistical properties, which is described in the following sub-sections.

\subsection{Cross-spectrum between modulation angle bins}

\begin{figure}[b]
\sidecaption
\includegraphics[width=\textwidth,trim=0.0cm 8.0cm 0.0cm 0.0cm,clip=true]{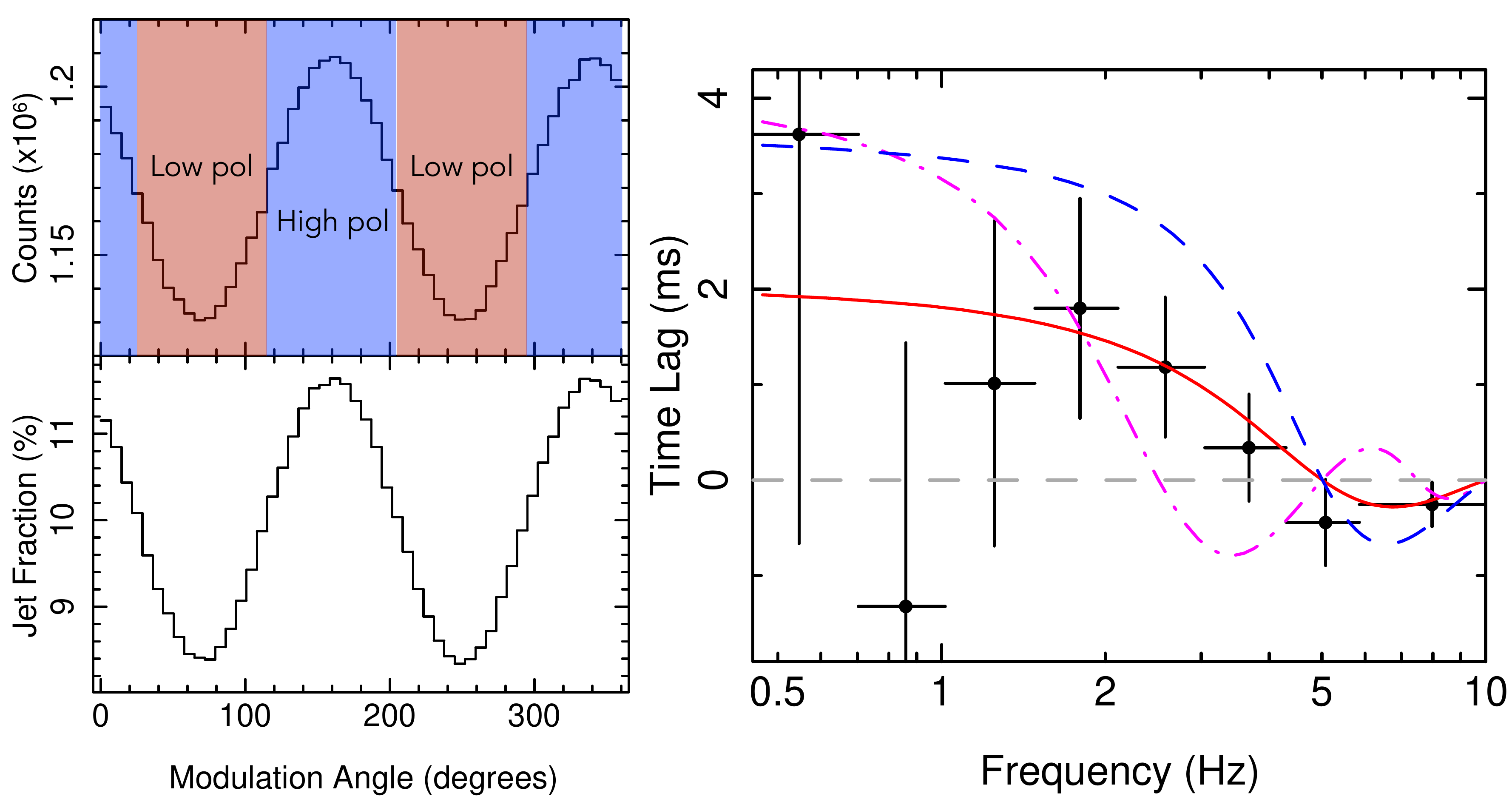}
\caption{Results of a simulation of accretion rate fluctuations propagating from a weakly polarized corona to a strongly polarized jet in some finite propagation time. \textit{a)}: Histogram of counts vs modulation angle recovered form the simulation (top). Since the jet is more highly polarized than the corona, a greater fraction of the total detected photons come from the jet for modulation angles close to the polarization angle (bottom). It is therefore useful to select `high' and `low' polarization light curves based on modulation angle (blue and red shading). \textit{b):} Time lag between high and low polarization light curves for the simulation (black points) and model (red solid line). Blue dashed and magenta dot-dashed lines represent alternative choices of parameters (details in the text).}
\label{fig:jetlag}       
\end{figure}

We seek a general method to detect and characterise stochastic variability in polarization properties, whether that be broad band noise or QPOs of any type. We need to understand the statistics well enough to be able to fit models, test goodness of fit and conduct null-hypothesis tests. A good starting point to consider a method that fits these criteria is to consider searching for time lags between a weakly polarized population of photons and a second highly polarized population of photons. A physical scenario for this was discussed in Section \ref{sec:propfluc}: mass accretion rate fluctuations propagating through a weakly polarized corona and eventually up a highly polarized jet. Even if these two populations of photons have the same spectrum, we can tell them apart if they have different polarization degrees because they will therefore have different modulation functions. Fig \ref{fig:jetlag} shows a simulated example. Here, we assume that the jet synchrotron radiation is $65\%$ polarized and contributes $10\%$ of the detected X-ray flux, whereas the corona radiation is $5\%$ polarized and contributes the remaining $90\%$ of the flux. Both components have the same polarization angle, $\psi=160^\circ$, the total count rate is $140$ c/s and the assumed exposure time is $420$ ks. The simulation considers $2^{23}$ time bins, each of duration $50$ ms. Panel a (top) shows the overall histogram of counts vs modulation angle for the simulation. Below this, the fraction of the counts that came from the jet is plotted as a function of modulation angle. We see that, because the jet emission is more highly polarized than the corona emission, a greater fraction of the counts are from the jet for modulation angles close to the overall polarization angle. This gives us a way to tell apart the two populations of photons. We can create a `high polarization' light curve, consisting of all the photons with modulation angle falling in the region colour-coded blue, and a `low polarization' light curve, consisting of all other photons (red shading). Since the high polarization light curve should have a larger jet contribution than the low polarization light curve, we expect the high polarization light curve to lag the low polarization light curve if there is indeed a propagation lag between the jet and corona.

In the simulation summarised by Fig \ref{fig:jetlag}, a $0.1$ s propagation lag is assumed between corona and jet, motivated by observations of the optical and IR signal lagging the X-ray signal (see Section \ref{sec:propfluc}). The power spectrum of corona variability is assumed to be a zero centered Lorenztian with HWHM = $6$ Hz, normalised to have an integrated variability amplitude of $20\%$, again motivated by observations. The black points in panel b show the time lag between simulated high and low polarization light curves, calculated by averaging the cross-spectrum between the two light curves over $2^{16}$ segments each of duration $6.4$ s and further employing geometrical frequency binning (the binning constant is 1.3). The red solid line represents the input model. We see that the high polarization light curve does indeed lag the low polarization light curve by $\sim 2$ ms. This is far less than the $0.1$ s propagation lag between corona and jet. The reason for this is dilution: the high polarization light curve does include more jet photons than the low polarization light curve does, but it is still primarily made up of corona photons. Likewise, the low polarization light curve still includes a contribution from jet photons. Also note that the time lag becomes negative at $\sim 5$ Hz. This is due to phase wrapping. For a simple scenario such as this whereby there are only two components with some time lag $\tau$ between them, the first phase wrap occurs at a frequency $\nu = 1/(2\tau)$ \cite{Uttley2014}, which is entirely consistent with the red line in panel b. Such frequency dependence gives us a way to disentangle dilution from the propagation lag itself. The blue dashed line is the same calculation but the jet fraction is now $20\%$, whereas the magenta dot-dashed line is for a jet fraction of $10\%$ but a propagation lag of $0.2$ s. We see that these two new lines both have a low frequency lag double that of the original calculation, but increasing the propagation lag halves the frequency of the first phase wrap.

Note that this simulation is very simple. For instance the corona and jet have the same polarization angle, although choosing different angles only complicates the picture rather than changing the fundamental point. Some of the chosen parameters are also on the optimistic side. A jet fraction of $10\%$ is rather high, a count rate of $140$ c/s represents a very bright state of e.g. Cyg X-1 observed with IXPE, and an exposure time of $420$ ks is very long. In particular, IXPE would not be able to observe such a bright source for such a long time without a break due to telemetry limitations. Still, such a long exposure could in principle be built up over a number of visits as long at the source spectrum stays relatively stable for a long period of time. These optimistic parameters are chosen to illustrate the concept that such an observation is in principle possible. In reality, a real detection analogous to the simulation in Fig \ref{fig:jetlag} may be beyond IXPE and instead require the larger collecting area of eXTP. In any case, the main importance of this discussion is to illustrate an important concept: we can select light curves based on modulation angle and calculate the cross-spectrum between them. Since we have been doing exactly the same thing for energy channels for decades, the statistics are very well understood.

\subsection{Modulation angle dependent cross-spectra}

As for the cross-spectrum between energy channels, the concept of making a cross-spectrum between two modulation angle bins can easily be extended to any number of modulation angle bins. We simply define $K$ modulation angle bins and extract a light curve for each. We can then take the Fourier transform of each of these light curves and calculate the cross-spectrum with some reference band light curve. This reference band light curve could, for example, be the total count rate from the polarimeter (summed over all modulation angle bins), or it could even be the count rate collected by a second detector pointed at the same source. The latter will be straight forward for eXTP, since the LAD will be on the same satellite. Bringing in a second detector would be much trickier for IXPE as it would involve using a second satellite (e.g. NICER), and time would be lost to Earth occultations and other drop-outs happening at different times for the two satellites. In any case, it is straight forward to calculate $K$ cross-spectra, and from that lag vs modulation angle spectra for given ranges in Fourier frequency. We can also easily calculate the modulus of the cross-spectra in order to plot the (correlated) variability amplitude in a given frequency range as a function of modulation angle. In all cases, the statistics are well understood and error bars can be calculated (see e.g. \cite{Ingram2019a}). For this reason, it is desirable to fit our models of polarization variability to these lag and variability amplitude vs modulation angle spectra.

So if we have a model that predicts Stokes parameters as a function of frequency, $Q(\nu)$ and $U(\nu)$, can we also calculate from these Stokes parameters a prediction for the cross-spectra for different modulation angles that can then be compared to data? For example, in Section \ref{sec:reverb} we derived expressions for a reverberation mapping model of time lags for I, Q and U (Equation \ref{eqn:tlag}). Let us refer to the cross-spectra we are hoping to derive as $C(E,\tilde{\psi},\nu)F^*(\nu)$, where $F(\nu)$ is the Fourier transform of the reference band light curve, $c(E,\tilde{\psi},t)$ is the count rate as a function of time and energy in a modulation angle of width $\Delta\tilde{\psi}$ and centred on $\tilde{\psi}$, and $C(E,\tilde{\psi},\nu)$ is its Fourier transform. We can write this `subject band' count rate as
\begin{equation}
    c(E,\tilde{\psi},t) = \frac{\Delta\tilde{\psi}}{2\pi} ~c(E,t)~ \bigg\{ 1 + \mu p(E,t) \cos\left[ 2 ( \psi(E,t) - \tilde{\psi} ) \right]  \bigg\},
\end{equation}
where $c(E,t)$ is the count rate summed over all modulation angle bins. Expanding out the cosine function with a trigonometric identity and subbing in the definitions of the Stokes parameters gives
\begin{equation}
    c(E,\tilde{\psi},t) = \frac{\Delta\tilde{\psi}}{2\pi} ~\bigg\{ c(E,t) + Q(E,t) \mu \cos(2 \tilde{\psi}) + U(E,t) \mu \sin( 2 \tilde{\psi} )  \bigg\}.
\end{equation}
It is therefore trivial to Fourier transform the above in order to express $C(E,\tilde{\psi},\nu)$ in terms of the Stokes parameters. If the reference band used is the total polarimeter count rate itself, the cross-spectrum integrated over all energies is
\begin{equation}
    C(\tilde{\psi},\nu) F^*(\nu)
    = \frac{\Delta\tilde{\psi}}{2\pi} ~\bigg\{ |C(\nu)|^2 + Q(\nu)C^*(\nu) \mu \cos(2 \tilde{\psi}) + U(\nu)C^*(\nu) \mu \sin( 2 \tilde{\psi} )  \bigg\}.
\end{equation}
It is therefore possible to fit any variability model that specifies $Q(\nu)$ and $U(\nu)$ to data via the modulation angle dependent cross-spectrum.

\subsection{Null hypothesis tests for polarization variability}

\begin{figure}[b]
\sidecaption
\includegraphics[width=\textwidth,trim=0.0cm 10.0cm 2.0cm 1.0cm,clip=true]{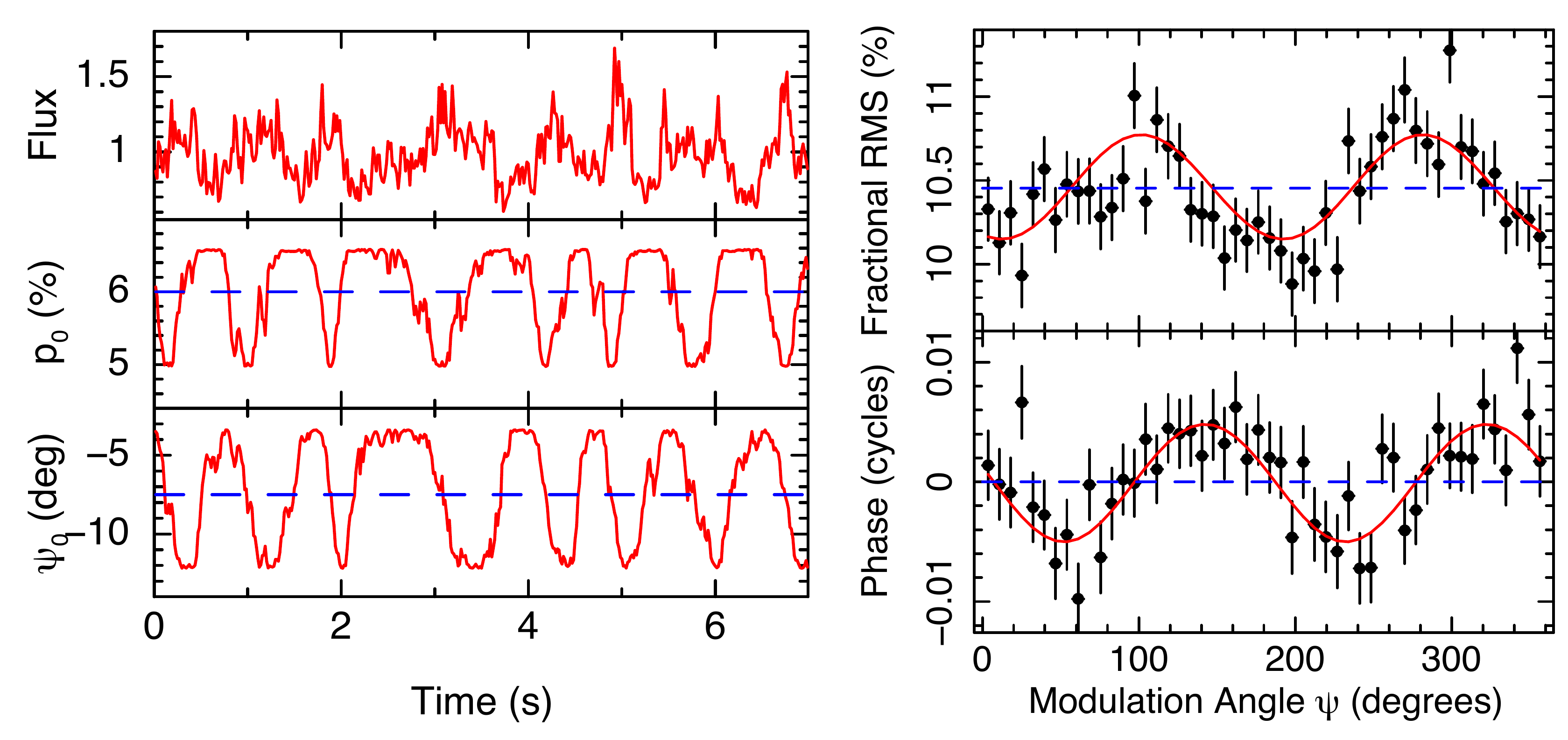}
\caption{\textit{Left:} Short segment of a simulation of X-ray flux (top), polarization degree (middle) and polarization angle (bottom) varying with time. The flux light curve includes a QPO and broadband noise. The QPO is generated from a precession model \cite{Ingram2015a}, and also includes a QPO in polarization degree and angle (red). Blue dashed lines represent the null-hypothesis, whereby the polarization properties are constant. \textit{Right:} Fractional rms (top) and phase lag (bottom) at the QPO fundamental frequency as a function of modulation angle for the full model (red solid line), the null-hypothesis model (blue dashed line), and a synthetic 200 ks IXPE observation based on the full models (black points). The sinusoidal shape of the full model is recovered by the synthetic data, and the null-hypothesis model can be ruled out.
}
\label{fig:poltest}       
\end{figure}

We now have a method to fit any specific model for $Q(\nu)$ and $U(\nu)$ to observational X-ray polarimeter data. However, what if we do not want to test a specific model for $Q$ and $U$ but instead test against a basic null hypothesis? Specifically, what if we want to test whether or not $p$ and $\psi$ are varying on a given variability timescale? Such a test would be extremely useful for QPOs, since
determining whether or not the polarization properties are modulated on the QPO period provides a clear diagnostic to differentiate between QPO models without the need for any fitting of specific model predictions. This sub-section describes how the modulation angle dependent cross-spectrum can be used for such a test.

Fig \ref{fig:poltest} demonstrates the method of Ingram \& Maccarone \cite{Ingram2017a}. The left panel shows synthetic time series of the flux, polarization degree and angle. The flux light curve includes broad band noise (generated with an `exponentiated' Timmer \& Konig simulation \cite{Uttley2005}) and a LF QPO. The LF QPO is generated from the `high inclination' precession model of \cite{Ingram2015a}, with the phase set to drift away from that of a pure sinusoid on a Gaussian random walk. The same precession model also predicts QPOs in polarization degree and angle, which we can see in the plot. The right hand panel shows the fractional rms variability amplitude and phase lag of the QPO fundamental as a function of modulation angle, both calculated from the modulation angle dependent cross-spectrum. The black points are simulated for a 200 ks IXPE observation with a mean count rate of 100 c/s and the red line is the input model. We see that variability in polarization degree and angle causes a sinusoidal dependence of both fractional rms and phase lag on modulation angle. If the polarization degree and angle were instead constant, both the fractional rms and phase lag would be independent of modulation angle (blue dashed lines). This enables a simple null-hypothesis test for the presence of polarization variability: testing whether or not the observed fractional rms and phase lag vs modulation angle are consistent with a straight line. In the simulation shown, the null-hypothesis (blue dashed lines) is strongly ruled out, meaning that this would constitute a $>3\sigma$ detection of a polarization QPO if these were real data.

\begin{figure}[b]
\sidecaption
\includegraphics[width=0.5\textwidth,trim=1.0cm 1.0cm 2.0cm 11.0cm,clip=true]{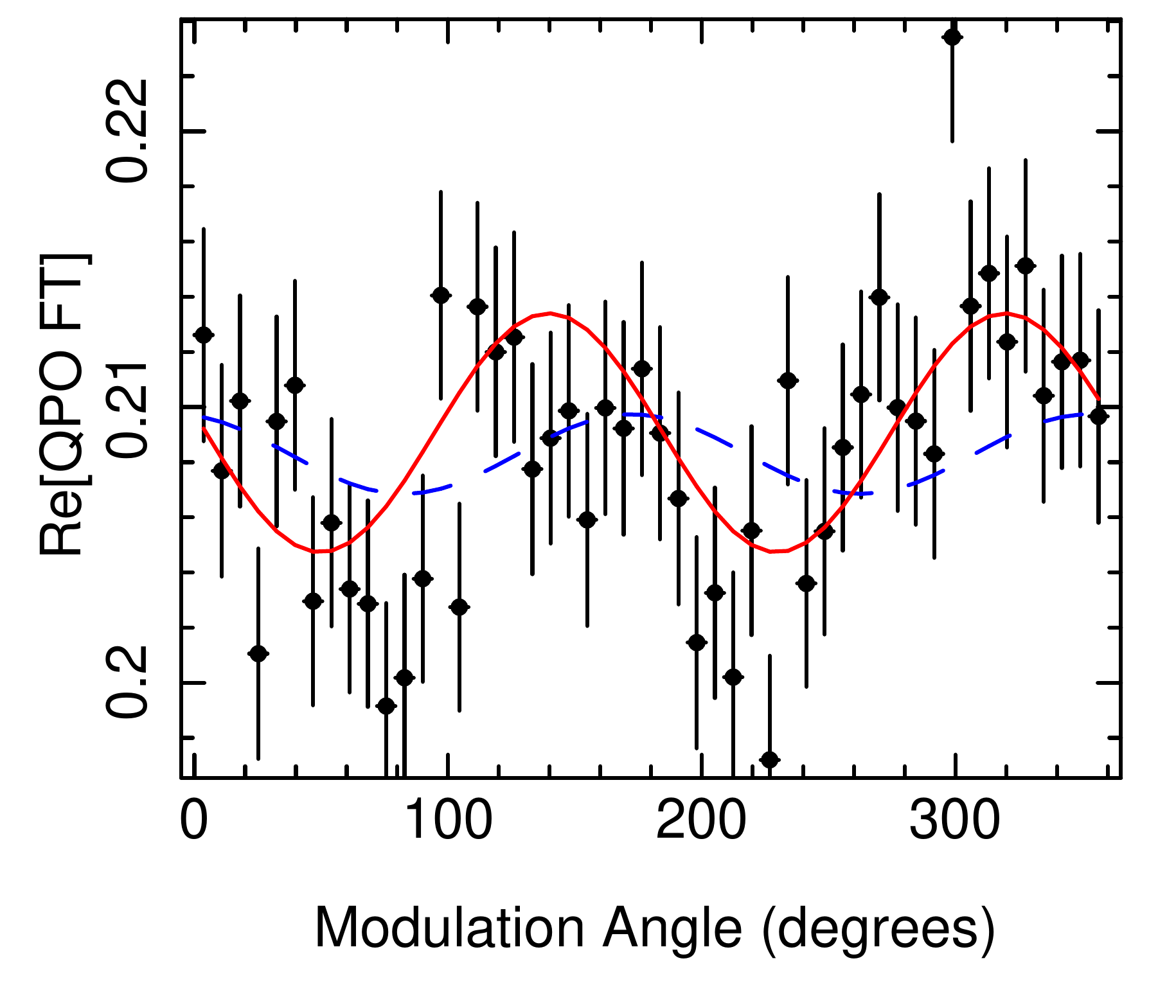}
\includegraphics[width=0.5\textwidth,trim=1.0cm 1.0cm 2.0cm 11.0cm,clip=true]{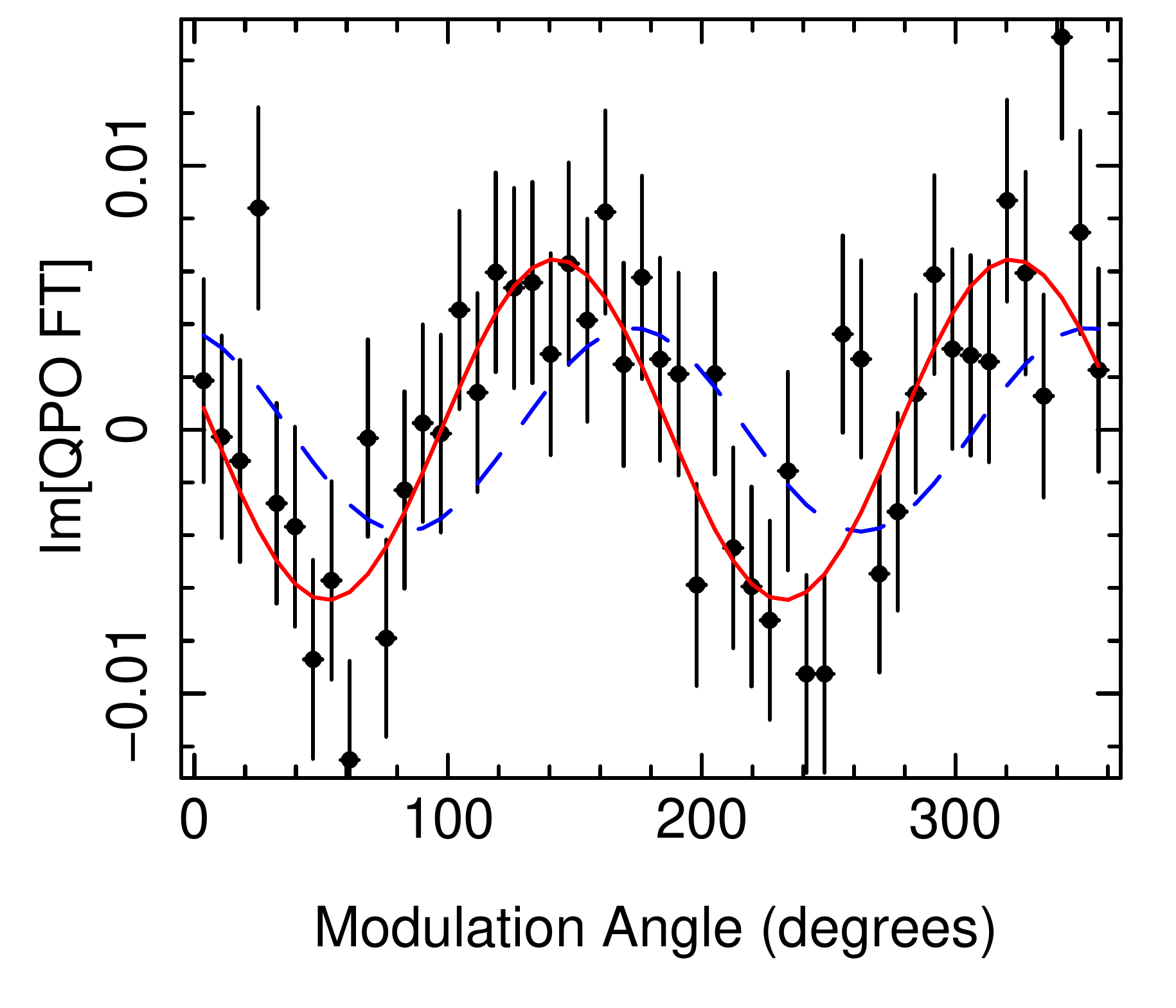}
\caption{Real (left) and imaginary (right) parts of the modulation angle dependent cross-spectrum at the QPO fundamental frequency, divided through by the amplitude of the QPO fundamental. This provides an estimate of the Fourier transform of the QPO signal. The red solid line represents the input model from Fig \ref{fig:poltest} and the black points represent a synthetic 200 ks IXPE observation. The blue dashed lines represent the best fit model in which the polarization angle does not vary with QPO phase. Since the synthetic data significantly prefer the full model, the null hypothesis of constant polarization angle can be ruled out.}
\label{fig:psitest}       
\end{figure}

The above null-hypothesis test is useful for ruling out a scenario whereby \textit{neither} the polarization degree \textit{nor} polarization angle are varying. It would be even more useful to have individual null-hypothesis tests for polarization degree and angle. Of the two, polarization angle is perhaps the most desirable property to test for a QPO modulation of, since quasi-periodic swings in the polarization angle are a particularly distinctive characteristic of precession -- in direct analogy to the radio polarization swings seen from pulsars. It turns out that a simple null-hypothesis test for polarization angle variability is indeed possible. If the polarization angle is constant, then the modulation angle dependent cross-spectrum becomes
\begin{equation}
 C(\tilde{\psi},\nu) C^*(\nu)
    = \frac{\Delta\tilde{\psi}}{2\pi} ~\bigg\{ |C(\nu)|^2 + \mu B(\nu)C^*(\nu) \cos[2 (\psi-\tilde{\psi})]\bigg\}.
    \label{eqn:psinull}
\end{equation}
Here, $B(\nu)$ is the Fourier transform of $b(t) = c(t) p(t)$, and is complex. Crucially, the values of ${\rm Re}[B(\nu)C^*(\nu)]$ and ${\rm Im}[B(\nu)C^*(\nu)]$ can simply be set as arbitrary parameters of our null-hypothesis model. The important insight from the above equation is that, if $\psi$ is not varying on a particular frequency, then the real and imaginary parts of $C(\tilde{\psi},\nu) C^*(\nu)$ will have a sinusoidal $\tilde{\psi}$ dependence that peaks at $\psi$. Therefore, if the sinusoidal dependence of the real and imaginary parts of the cross-spectra reach their peak significantly away from the mean measured polarization angle, we can conclude that $\psi$ must be varying on that frequency.

Fig \ref{fig:psitest} shows an example. This is the same simulation as is shown in Fig \ref{fig:poltest}, but now the real and imaginary parts of the cross-spectrum are plotted instead of the amplitude and phase. Again, the black points represent a synthetic 200 ks IXPE observation and the red lines the input model. The blue dashed lines represent the best fitting constant $\psi$ null-hypothesis model; i.e. equation (\ref{eqn:psinull}) with ${\rm Re}[B(\nu)]$ and ${\rm Im}[B(\nu)]$ left as free parameters in a least squares fit to the synthetic data. The red lines describe the data significantly better than the blue dashed lines, meaning that it is possible to conclude with $> 3 \sigma$ confidence that the polarization angle is modulated on the QPO fundamental in this synthetic data set. There is currently no such null-hypothesis test in the literature to test for a modulation in polarization degree.

These null-hypothesis tests can also be applied to HF QPOs \cite{DeRosa2019}, but beyond that they are not married at all to the presence of QPOs. We can use these tests to determine whether there is variability in the polarization properties in any Fourier frequency range, regardless of whether or not that range contains a QPO peak.

\subsection{Technical challenges}

A clear advantage of the modulation angle dependent cross-spectrum is that it has its heritage in the already well-used energy dependent cross-spectrum, and therefore is well understood. There are still, however, technical challenges and pitfalls to overcome. The first has a very simple workaround, but is nonetheless important to discuss: the effects of background. The background is constant in time and un-polarized, whereas the source is (hopefully) polarized and with a variable count rate. The total signal is therefore a combination of two population of photons. During epochs when the source count rate is at its highest, the dilution from the unpolarized background is low and therefore the polarization of the combined source plus background signal is at a peak. When the source count rate is at its lowest, dilution from the background is maximal and the polarization therefore reaches a minimum. Specifically, the polarization of the combined signal and background, $p_{\rm sb}(t)$, is related to the polarization of the source, $p(t)$, as
\begin{equation}
    p_{\rm sb}(t) = \frac{s(t)}{\langle b \rangle + s(t)}~p(t).
\end{equation}
It is clear from the above that even if $p$ is constant in time, $p_{\rm sb}$ is variable because $s$ is variable. Such spurious polarization variability will only be a small effect for IXPE which has a low background ($b \ll s$), and can easily be avoided completely simply by using background subtracted light curves. For IXPE, it is simple to estimate the background as it is an imaging instrument.

\begin{figure}[b]
\sidecaption
\includegraphics[width=\textwidth,trim=20.0cm 11.0cm 29.0cm 12.0cm,clip=true]{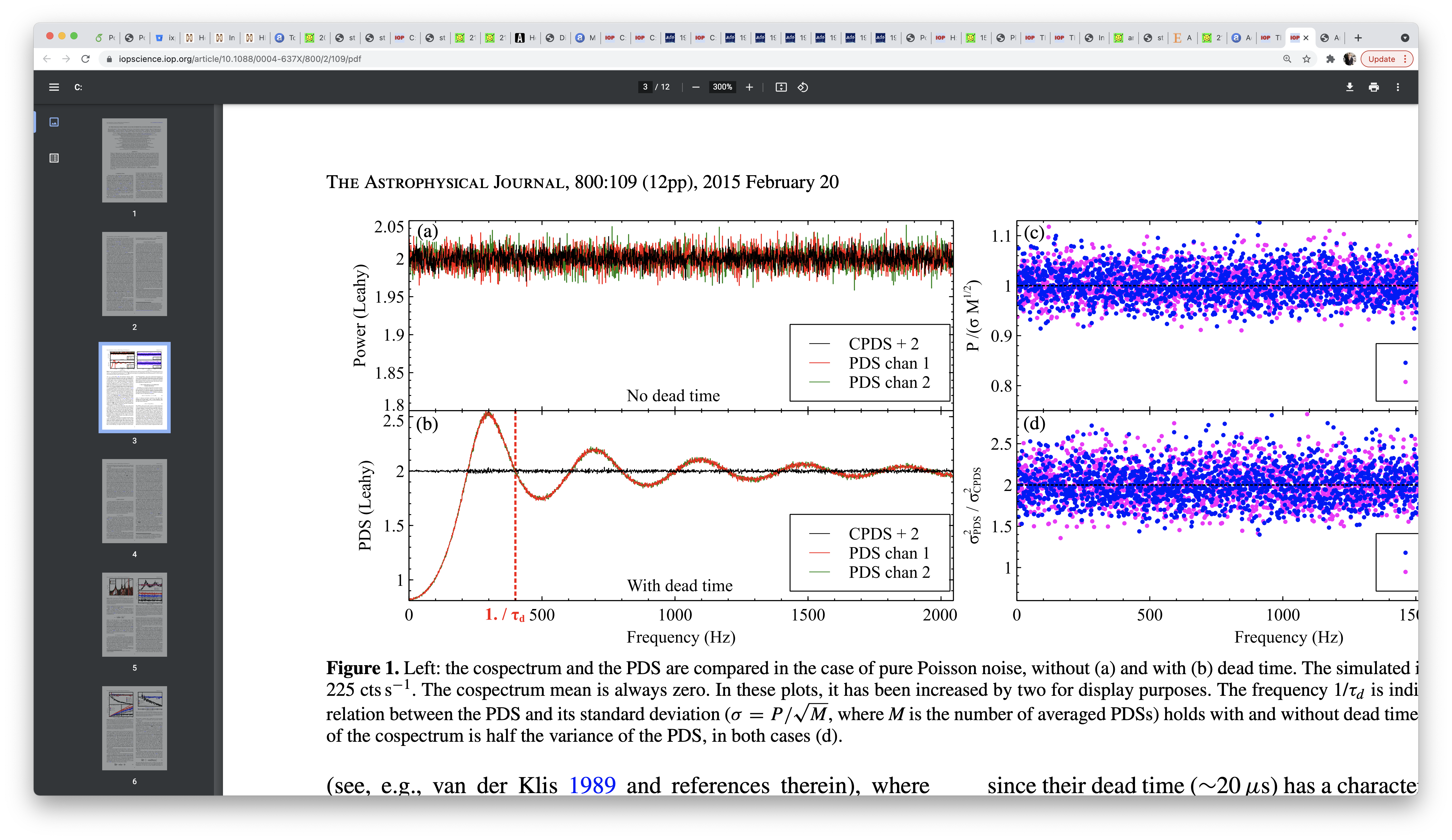}
\caption{Results of a simulation to demonstrate the effects of deadtime on the power spectrum. A light curve is simulated assuming a constant source flux and a mean count rate of $225$ c/s, meaning that all variability in the light curve is due to counting statistics. Red and green represent the resulting power spectrum of two independent detectors. The top plot is in the absence of detector deadtime, and the bottom is in the presence of a deadtime of $\tau_d=2.5$ ms. Deadtime imprints a distinctive sinc function on the Poisson noise contribution. The power spectrum are in Leahy normalisation, ensuring that the Poisson noise level is 2 in the absence of deadtime. Black points represent the co-spectrum between the two detectors, plus a constant of 2. We see that deadtime does not print a sinc function on to the co-spectrum. Figure reproduced from \cite{Bachetti2015}.}
\label{fig:deadtime}       
\end{figure}

A much more involved challenge is presented by detector deadtime. This is the waiting time, $\tau_d$, after a photon has been detected during which the detector is not sensitive to any more photons. Therefore, the detector sensitivity depends on the incident count rate itself: the larger the count rate, the greater the fraction of incident photons that are missed. Specifically, the detected count rate $c_{\rm det}$ relates to the incident count rate $c_{\rm in}$ as $c_{\rm det} = c_{\rm in} / ( 1 + \tau_d c_{\rm in} )$.
For a variable incident count rate, the detector sensitivity is therefore time-dependent, causing a number of difficulties for variability analyses.
The IXPE deadtime, $\tau_d \sim 1.2$ ms, is reasonably large and so the effects are significant. One effect of deadtime is on the shape of the Poisson noise in the power spectrum.
Fig \ref{fig:deadtime} demonstrates this with a simulation \cite{Bachetti2015}. This simulation is in the context of NuSTAR, which has two detectors (FPMA and FPMB) and a deadtime of $\tau_d \approx 2.5$ ms.
The source is set to be intrinsically constant in flux and all variability in the incident count rate results exclusively from Poisson counting statistics. The figure shows the resulting power spectrum measured for the two detectors (red and green) when deadtime is set to $\tau_d=0$ (top) and when it is set to $\tau_d=2.5$ ms (bottom).
The constant Poisson noise level seen in the absence of deadtime (top) is in stark contrast to the sinc function imprinted on to the power spectrum by deadtime (bottom). Another problematic effect of deadtime is its effect on phase lags. The detector sensitivity is varying in anti-phase with the total incident count rate, which leads to $\sim \pi$ radian phase lags between different energy channels or modulation bins from a given detector.

In principle, the effects of deadtime can be corrected for with a detailed model of the instrument. For instance, RXTE had a working deadtime model that could be used to accurately reproduce the Poisson noise contribution to the power spectrum \cite{Zhang1995}. In practice, however, deriving a working deadtime model for an imaging instrument is very difficult as different categories of events have different associated deadtime values. Modelling the effects of deadtime on the phase lags is even more difficult. An elegant solution is used for NuSTAR, which makes use of the two detectors being independent of one another \cite{Bachetti2015}. Instead of calculating the power spectra of the FPMA and FPMB signals individually, we can instead calculate the real part of the cross-spectrum between the FPMA and FPMB signals -- known as the co-spectrum. Poisson noise averages out to zero in the co-spectrum because the counting noise in the FPMA is uncorrelated with that in the FPMB. The black points in Fig \ref{fig:deadtime} show the co-spectrum of the simulation plus a constant. We see that no sinc function is imprinted onto the co-spectrum even for $\tau_d=2.5$ ms, circumventing one of the issues caused by deadtime. A similar work-around can be used for phase lags: deadtime is circumvented if we extract our subject band light curves from the FPMA and our reference band light curve from the FPMB (or vice versa) \cite{Ingram2016}. Both of these tricks can also be used for IXPE, which has three independent gas pixel detectors (GPDs). We can therefore calculate the co-spectrum between, say the GPD1 light curve and another light curve made by summing the GPD2 and GPD3 photons. To calculate phase lags, we can use e.g. GPD3 for the reference band and GPD1+GPD2 for the subject bands. It of course doesn't matter \textit{which} detectors are used for which purpose, but signal to noise will be maximised by using two detectors for the subject bands and one detector for the reference band. This is because, for a given total count rate, the maximum signal to noise is achieved when the count rate of the subject band is equal to that of the reference band \cite{Ingram2017a}. 

Although correlating the signals from independent detectors provides a simple and elegant solution to the deadtime problem, it is not optimal in terms of signal to noise. The optimal solution in terms of signal to noise ratio, would be to successfully model the effects of deadtime. A simple way to account for our imperfect understanding of the detectors is to model the Poisson noise contribution with a sinc function, but leave the parameters of that sinc function free in a least squared fit to the high frequency power spectrum (which is dominated by Poisson noise) \cite{Bult2017}. This approach is attractive in its simplicity, but it does not address the larger problem associated with the phase lags. A more sophisticated solution is to use simulation based inference techniques. These have recently been demonstrated to be able to extract observables out of NuSTAR power spectra in an unbiased manner \cite{Huppenkothen2021}, and could in principle also be applied to the phase lag problem.

A final technical challenge results from IXPE observations being necessarily long in duration (in order to collect enough photons), whereas X-ray binaries can change state reasonably quickly. It is therefore possible that the spectrum and variability properties may change significantly during a long-look observation. This problem is accentuated by telemetry limitations. IXPE telemeters a lot of information for each detected photon. The time-dependent voltage on every pixel is sent down to Earth so that the modulation angle associated with each photon is calculated on the ground instead of on the satellite. The significant advantage of this is that different algorithms can be applied to the same raw data retrospectively. If, for example, someone comes up with a better way to extract modulation angles in 5 years time, they will be able to apply their technique to IXPE archival data instead of needing to wait for another X-ray polarimeter to fly with their software on board. Coupled with the relatively low bandwidth of telemetry available to IXPE, however, this leads to bright X-ray binaries generating data at a faster rate than IXPE can telemeter it to Earth. The on board buffer therefore fills up until a point where it is not possible to record any new data without overwriting the buffer. Long observations of bright sources therefore need to be split up and interspersed with observations of dim sources, during which the stored information from the previous observation can be transmitted. The upshot is that the elapsed time of a long-look observation can be very large indeed compared with the timescale on which the spectrum and timing properties of X-ray binaries typically change. One partial solution for QPOs in particular is to track the QPO frequency across many short time intervals and stack based on this instantaneous QPO frequency instead of the average frequency.

\section{Conclusions}

X-ray polarimetry-timing will provide a novel way to study compact objects. This Chapter has summarised scenarios in which we may theoretically expect the polarization degree and angle of X-ray radiation from NSs and BHs to vary on short timescales, and how we might exploit this variability in order to learn about fundamental physics and the objects themselves. Since the recent launch of IXPE, there is finally an X-ray polarimeter with fast timing capability, and planned missions in the near future such as eXTP will provide even better data. Table \ref{tab:1} summarises the scenarios discussed in this Chapter, and provides estimates of the first X-ray polarimetry mission that will be sensitive enough to make a detection of each predicted effect. Still, observing variability in X-ray polarization and angle is technically difficult due to the statistical nature of how X-ray polarization is detected. This Chapter has summarised the techniques that can be used, enabling fitting of models for polarization variability to data and simple null-hypothesis tests for polarization variability. These techniques can soon be applied to IXPE data to search for, for example, QPOs in the polarization degree and angle of X-ray binaries.

%
\begin{table}[!t]
\caption{Theoretical predictions and polarimetry-timing techniques discussed in this chapter with a section reference and a prediction of the first X-ray polarimetry mission that will be able to achieve the necessary sensitivity. Typically, where `future' missions are predicted to be required, the required characteristics are high effective area, large modulation factor and high count rate capability (e.g. small deadtime, not susceptible to pileup etc).}
\vspace{2mm}
\label{tab:1}       
%
%
\begin{tabular}{p{6cm}p{2cm}p{3cm}}
\hline\noalign{\smallskip}
Prediction / Technique  & Section & Mission  \\
\noalign{\smallskip}\svhline\noalign{\smallskip}
Pulse phase-resolved polarimetry & \ref{sec:pulsars} & IXPE \\
Vacuum birefringence & \ref{sec:pulsars} & IXPE \\
Magnetic condensation & \ref{sec:pulsars} & IXPE \\
Disc-corona propagation & \ref{sec:propfluc} & eXTP/future \\
Corona-jet propagation & \ref{sec:propfluc} & eXTP/future \\
Polarized reverberation mapping & \ref{sec:reverb} & eXTP/future \\
Polarization LF QPOs & \ref{sec:QPOs} & IXPE/eXTP \\
Polarization HF QPOs & \ref{sec:QPOs} & eXTP \\
Blazar variability & \ref{sec:blazars} & IXPE/eXTP \\
\noalign{\smallskip}\hline\noalign{\smallskip}
\end{tabular}
\end{table}

\bibliographystyle{spphys}
\bibliography{biblio}

\end{document}